# The Impact of Traceability on Software Maintenance and Evolution: A Mapping Study

Fangchao Tian [1,2], Tianlu Wang [1], Peng Liang [1*], Chong Wang [1], Arif Ali Khan [3], Muhammad Ali Babar [2]

[1] School of Computer Science, Wuhan University, Wuhan, China
{tianfangchao, wangtianlu, liangp, cwang}@whu.edu.cn

[2] School of Computer Science, The University of Adelaide, Adelaide, Australia
ali.babar@adelaide.edu.au

[3] Faculty of Information Technology, University of Jyväskylä, Finland
arifaan_beghian@yahoo.com

*Abstract*—Software traceability plays a critical role in software maintenance and evolution. We conducted a systematic mapping study with six research questions to understand the benefits, costs, and challenges of using traceability in maintenance and evolution. We systematically selected, analyzed, and synthesized 63 studies published between January 2000 and May 2020, and the results show that: traceability supports 11 maintenance and evolution activities, among which *change management* is the most frequently supported activity; strong empirical evidence from industry is needed to validate the impact of traceability on maintenance and evolution; easing the process of change management is the main benefit of deploying traceability practices; establishing and maintaining traceability links is the main cost of deploying traceability practices; 13 approaches and 32 tools that support traceability in maintenance and evolution were identified; *improving the quality of traceability links*, *the performance of using traceability approaches and tools* are the main traceability challenges in maintenance and evolution. The findings of this study provide a comprehensive understanding of deploying traceability practices in software maintenance and evolution phase, and can be used by researchers for future directions and practitioners for making informed decisions while using traceability in maintenance and evolution.

1 INTRODUCTION

Software traceability refers to the ability that traces various types of artefacts (e.g., requirements specifications, architecture documents, test models, source code) and establishes the links between them during the software development process. Software traceability is recognized as an important factor to support various software development activities [37], and software maintenance and evolution is known to incur significant cost and slow down the speed of software implementation [1].

A large number of studies show that traceability plays a significant role in software maintenance and evolution. The traces between various software artefacts (e.g., requirements, design, and code) can facilitate software maintenance and evolution [3][4][5]. For example, a recent mapping study conducted by Charalampidou *et al.* [45] revealed that using traceability can benefit quality attributes related to maintainability, such as modifiability, analyzability, and testability. The experimental study conducted by Mäder and Egyed [3] shows that the subjects supported with traceability performed faster on a given maintenance task and created more correct solutions to the task. Such observations from the literature indicate that traceability can not only save the effort of maintenance activities (e.g., change management), but can also improve the quality of software maintenance as the traces between various artefacts can help developers better understand the system during software maintenance [3].

Deploying traceability in software maintenance and evolution phase could increase the overall cost of development. It is a challenging and time-consuming job to develop traces [12]. Moreover, as the system evolves, it can be even more difficult to precisely maintain the traces [13]. In safety-critical systems, traceability often delays the development activities by adding additional overhead to the maintenance phase [10]. Consequently, a majority of the software development organizations hesitate to adopt the traceability practices as the impact (e.g., costs and benefits) of traceability on software maintenance and evolution phase is not clear. The uncertain cost-benefit ratio of traceability is still a critical issue for practitioners who are struggling to understand the significance of traceability [31]. Therefore, it is important to systematically investigate the impact of traceability on maintenance and evolution (e.g., the costs and benefits of using traceability), which could provide a roadmap to both researchers and practitioners for exploring and employing traceability practices in software maintenance and evolution.

---

* Corresponding author

Systematic Mapping Study (SMS) is a type of secondary study that provides a systematic overview of a particular research topic, by analyzing the available primary studies and answering the questions on a broad topic [18][19]. Previously, various secondary studies (see Section 7) were conducted in the area of software traceability (e.g., requirement traceability [3] and architecture traceability [20]), but as revealed by the SMS on software traceability [45], there is lack of an in-depth exploration of the primary studies that provide empirical evidence on the relation between traceability and software maintenance and evolution. To fill this gap, we conducted an SMS to explore the existing primary studies that present the existing approaches, tools, and empirical evidence about the association between traceability and software maintenance and evolution [18]. Similar to systematic mapping study, Systematic Literature Review (SLR) is an approach to conduct secondary studies by identifying, evaluating, and interpreting all available primary studies relevant to a particular topic [18]. We selected the SMS approach because the impact of traceability on software maintenance and evolution is a broad topic which covers various aspects. However, we stressed upon in this SMS the empirical evidence of the primary studies about the traceability impact on maintenance and evolution activities, costs and benefits of deploying traceability practices approaches and tools that support the use of traceability during the maintenance and evolution phase, and challenges of deploying traceability practices.

The rest of this paper is organized as follow. Section 2 presents the definition and explanation of software traceability, maintenance and evolution activities. Section 3 defines the mapping study process including the research questions and the execution process of this mapping study. Section 4 provides the results to the research questions. Section 5 further discusses the study results. Section 6 presents the threats to validity. Section 7 compares the related secondary studies with this study, and Section 8 concludes this mapping study.

2 RESEARCH CONTEXT

Before describing the mapping study process, we provide the definitions and explanations of two key concepts, i.e., "*software traceability*" and "*software maintenance and evolution*". The given definitions will clarify the scope and domain of the study.

2.1 *Software Traceability*

In the IEEE standard glossary of software engineering terminology [2], software traceability is defined as "*the degree to which a relationship can be established between two or more products of the development process*", which is a critical internal quality attribute of software systems. Software traceability also refers to the ability to create the links between various types of software artefacts [37]. In general, traceability can be categorized into vertical and horizontal traceability, depending on whether traceability links associate artefacts at the same level of abstraction (i.e., horizontal traceability, for example, traceability between requirements) or artefacts at different levels of abstraction (i.e., vertical traceability, for example, traceability from requirements to design) [38]. Traceability among various software artefacts (e.g., requirements, design, and code) can support many activities in the development of software systems. More specifically, software traceability can ease the analysis process of changes in software maintenance, evolution, and reuse and testing of software components [37]. In recent years, studies have been conducted in various contexts of software traceability, such as traceability approaches [39], traceability deployment and maintenance of traceability links [40][41].

2.2 *Software Maintenance and Evolution*

Software maintenance and evolution is a critical software development phase. The IEEE Standard 1219 [26] define software maintenance as "*the modification of a software product after delivery to correct faults, to improve performance or other attributes, or to adapt the product to a modified environment*". Software maintenance focuses on fixing bugs to prevent software failing and preserve the intended functionalities [30]. The term software evolution lacks a standard definition and is often used as a substitute for software maintenance [1][30]. In this study, we do not distinguish maintenance and evolution strictly and use both to refer to the update or modification in software systems. Software maintenance and evolution are recognized as expensive activities due to their significant impact on cost and time [1]. Both are inevitable activities as most of the software systems are likely to be changed and improved during development process or after release.

## 3 MAPPING STUDY PROCESS

### 3.1 Research Questions

The goal of this mapping study defined using the Goal-Question-Metric (GQM) approach [11] is: *to analyze* the impact of traceability on maintenance and evolution *for the purpose of* exploration and analysis *with respect to* the activities, empirical evidence, traceability approaches, tools, benefits, costs, and challenges *from the point of view of* researchers and practitioners *in the context of* software development. The reasons that we considered these aspects are that activities are the basic unit of software maintenance and evolution, traceability approaches and tools provide the ways of using traceability, and benefits, costs, and challenges are representations of the impact of traceability.

Six Research Questions (RQs) as shown in TABLE I were developed to achieve the study goal. The results and analysis of the given RQs can be directly associated with the goal of the study: maintenance and evolution activities supported by traceability (RQ1), empirical evidence that supports the use of traceability in maintenance and evolution activities (RQ2), the benefits, costs, and challenges of using traceability in maintenance and evolution activities (RQ3, RQ4, RQ6), and the approaches and tools that support traceability in maintenance and evolution activities (RQ5).

TABLE I. RESEARCH QUESTIONS AND THEIR RATIONALE

| Research Question | Rationale |
|---|---|
| **RQ1**: What software maintenance and evolution activities can be supported by traceability? | Traceability can support various software maintenance and evolution activities. The answer of this RQ reveals how traceability can facilitate the maintenance and evolution phase. |
| **RQ2**: How much evidence has been reported to support the use of traceability in software maintenance and evolution activities? | To gain an understanding of the validity and assess the quality of the primary studies, this RQ investigates the evidence level in each study concerning the impact of traceability on maintenance and evolution activities. |
| **RQ3**: What would be the benefits of using traceability practices during the software maintenance and evolution phase? | Traceability brings various benefits while deploying its practices during the software maintenance and evolution phase. The answer of this RQ could assist practitioners to make better decisions and improve the process of using the traceability practices. |
| **RQ4**: What would be the costs of using traceability practices during the software maintenance and evolution phase? | The answer of this RQ will provide an overview of costs concerning traceability, which can help practitioners properly manage the traceability budget and make cost-effective decisions. |
| **RQ5**: What approaches and tools have been used to support traceability during the software maintenance and evolution phase? | The answer of this RQ can provide practitioners with an overview of the readily available approaches and tools that they can use to employ the traceability practices in maintenance and evolution. |
| **RQ6**: What are the challenges of using traceability in the maintenance and evolution of software systems? | Using traceability in software maintenance and evolution has many challenges. The answer of this RQ will provide researchers with promising directions to be tackled for deploying the traceability practices. |

### 3.2 Mapping Study Execution

We followed the guidelines proposed by Kitchenham and Charters [19] to conduct this SMS, which consists of five phases, i.e., study search, study selection, snowballing, data extraction, and data synthesis. These five phases are detailed below.

#### 3.2.1 Study Search

Petersen *et al.* [19] identified three study search methods used in the existing SMSs and SLRs. We only used the two complementary search approaches in this SMS, i.e., automatic database search and snowballing [9]. The reason that we did not employ manual search is that traceability as a common topic in software engineering can be published in diverse venues. Automatic database search is used to retrieve the relevant studies from the electronic databases using search strings; snowballing is used to collect those relevant studies that are missed during the automatic search process. The details of using the two search approaches in this SMS are provided in the following sections.

*a) Search Scope*

The search scope of this study includes the selected databases and timeframe for retrieving the relevant studies. The following describes how these two elements are determined.

**Timeframe**: We only considered the primary studies published from January 2000 to May 2020. The starting year is justified by considering the roadmap paper of software maintenance and evolution that was published in 2000 and formulated the promising research directions in this regard in the last 20 years [1]. The end date is the time when we started this SMS, i.e., May 2020.

**Electronic databases**: We finally selected seven core Electronic Databases (EDs) (see TABLE II) based on the guideline provided by Chen *et al.* [8]. Google Scholar was not included in this study, as it would produce a significant number of irrelevant studies and overlap with the studies returned from other databases. Note that the search terms were only matched with paper titles and abstracts in the databases ED1, ED3, ED5, and ED6 due to the search limitation of these databases.

TABLE II. ELECTRONIC DATABASES FOR THE AUTOMATIC SEARCH

| No. | Electronic database | Search terms used in |
|---|---|---|
| ED1 | ACM Digital Library | Paper title, abstract |
| ED2 | IEEE Xplore | Paper title, keywords, abstract |
| ED3 | Springer Link | Paper title, abstract |
| ED4 | Science Direct | Paper title, keywords, abstract |
| ED5 | Wiley InterScience | Paper title, abstract |
| ED6 | EI Compendex | Paper title, abstract |
| ED7 | ISI Web of Science | Paper title, keywords, abstract |

*b) Search Strategy*

a. The search strategy influences the effort required to conduct the search process and the completeness of search. In this SMS, we used two most frequently applied methods for developing the search terms [19]. We initially defined the search terms based on the research topics and research questions (i.e., traceability and software maintenance and evolution) in the existing relevant studies (in Section 7). Moreover, PICO (Population, Intervention, Comparison, Outcomes) criteria [18] were used to develop the search terms. The population in this study is software maintenance and evolution, and the intervention refers to traceability. Only P and I were considered to develop the search terms, because traceability approaches are not compared with other approaches and the outcomes of using traceability are not limited in this SMS. We included various search terms that are highly relevant to traceability, maintenance and evolution. The word "software" was used to limit the scope of the search process and decrease the number of irrelevant studies (noise).

b. The selected electronic databases were explored for pilot search using the combinations of different search terms. Two search strings were used in this pilot, i.e., ("*maintain*" **AND** "*trace*") and (("*maintainability*" **OR** "*maintaining*") **AND** "*trace*"). We noticed that the search engines of the selected databases perform differently using the search terms, e.g., using the term "*maintain*" in IEEE Explore did not return the studies that were searched using the term "*maintainability*" or "*maintaining*". The similar situation also happened while the terms "*trace*" and "*traces*" were executed. Therefore, we included all these terms in the final construction of search strings to avoid missing potentially relevant studies. Note that we did not include specific terms of maintenance and evolution activities (e.g., "*bug fix*" and "*refactoring*") in the search query due to two reasons: part of the goal of this SMS is to identify the potential maintenance and evolution activities that can be supported by using traceability (i.e., RQ1), therefore we cannot have a relatively comprehensive list of those activities before conducting this SMS; and including the specific maintenance and evolution activity terms in the search query may lead to the situation that the search results are biased to those activities.

Both Boolean "**OR**" and "**AND**" operators were used to concatenate the terms for population and intervention. For example, if the search engine supports logical **OR** and **AND** operators, the search query could be composed as follow:

(*trace* **OR** *traces* **OR** *tracing* **OR** *traceability*) **AND** (*maintain* **OR** *maintaining* **OR** *maintenance* **OR** *maintainability* **OR** *evolve* **OR** *evolving* **OR** *evolution* **OR** *evolvability*) **AND** (*software*)

3.2.2 *Study Selection*

This section describes the inclusion/exclusion criteria for study selection as well as the details of the selection in three rounds.

**Selection Criteria**: The inclusion and exclusion criteria were developed by following the guidelines in [19] to refine the selection process of the identified studies (see TABLE III). The selection of a primary study was based on the title, abstract, and full text against the inclusion and exclusion criteria. Moreover, we recorded the key terms that lead to the inclusion/exclusion of a specific study, and those terms were further used for discussion during the consensus meetings and reassessment process.

TABLE III. INCLUSION AND EXCLUSION CRITERIA

| No. | Inclusion criteria |
|---|---|
| I1 | A paper that focuses on using traceability practices in software maintenance and evolution |
| I2 | A paper that is peer-reviewed and available in full text |
| **No.** | **Exclusion criteria** |
| E1 | If two papers publish the same work in different venues (e.g., conference and journal), the study with limited details is excluded. |
| E2 | Grey literature (i.e., technical report, work in progress) is excluded. |
| E3 | A paper not written in English is excluded. |
| E4 | A paper that is duplicated with an included paper. |

**Selection Process**: The search strategy defined in Section 3.2.1 was used to identify the relevant primary studies. The returned number of studies in each round is provided in Section 4.1.1. The selection process includes the following three core rounds:

a. **First round of selection**. The first author reviewed the titles of the studies retrieved from the digital databases using the selection criteria discussed in Section 3.2.2. Those studies were retained for the second round for which the first author had not made any decision.

b. **Second round of selection**. Both the first and second authors independently conducted a pilot selection by reading the abstracts of randomly selected 50 articles retained during the first round. The selected studies were put forward for discussion during the consensus meetings in order to resolve the disagreements between the first and second authors. Then, they selected papers independently by reading the abstracts of the remaining studies from the first round. Those studies are kept for the final round which were hard to decide based on their abstracts.

c. **Final round of selection**. The first and second authors independently conducted a pilot selection with randomly selected 20 articles retained during the second round. Any disagreements were discussed and resolved among all the authors by holding consensus meetings. Then, the first two authors conducted a full-length reading of the remaining papers. For uncertain papers, their final decisions were made based on the discussion of all the authors.

3.2.3 *Snowballing*

The snowballing approach was used to get those relevant studies that were possibly missed during the automatic database search. We adopted backward snowballing [9] in this SMS, which is also used in many other SMSs, such as [19]. Snowballing is an iterative process, where we checked the references lists of the studies selected from the final round of automatic database search (in Section 3.2.2), then the newly selected papers of the last iteration were checked in the subsequent iterations. The iterative process was completed when there were no newly selected papers. Each iteration followed the three rounds of selection process described in Section 3.2.2, i.e., based on the title, abstract, and full content. The papers selected during the snowballing process were included in the final list of the selected studies.

*3.2.4 Data Extraction*

Fifteen data items have been defined to be extracted from the primary studies in order to provide demographic information and answer the Research Questions (RQs) (see TABLE IV). The eight data items (D1-D8) are used to extract the demographic details of the primary studies and the remaining data items (D9-D15) are used to answer the six RQs. Both D9 and D11 are used to answer RQ3, which focuses on investigating the benefits (D11) brought by using traceability practices in various maintenance and evolution activities (D9).

The description of the data items and their relevant RQs are presented in TABLE IV. Interpersonal biases and misunderstanding regarding the data items were minimized by conducting pilot data extraction, which was followed by both the first and second authors. They randomly selected 10 primary studies and extracted the data according to the data items in TABLE IV. In the formal data extraction, the first two authors extracted the data from the remaining studies independently, which was then merged by the first two authors and further reviewed by the third and fourth authors. The conflicts, disagreements, and uncertainties in the data extraction process were resolved during the regular consensus building discussions during which all the authors provided their feedback and suggestions. The outputs of the data extraction process were formally recorded on separate Excel spreadsheets for further analysis and synthesis.

TABLE IV. DATA ITEMS EXTRACTED FROM THE PRIMARY STUDIES WITH THE RELEVANT RQS

| # | Data item name | Description | Relevant RQ |
|---|---|---|---|
| D1 | Index | Paper ID | Overview |
| D2 | Title | Paper title | Overview |
| D3 | Author list | Name of all the authors | Overview |
| D4 | Year | Year of publication | Overview |
| D5 | Venue | Name of the venue where the study is published | Overview |
| D6 | Publication type | Journal, conference, workshop, or book chapter | Overview |
| D7 | Author type | Researcher, practitioner or both | Overview |
| D8 | Research type | The research type of paper | Overview |
| D9 | Maintenance and evolution activities | Maintenance and evolution activities that can be supported by traceability | RQ1, RQ3 |
| D10 | Evidence level | The evidence level of the impact of traceability on software maintenance and evolution activities | RQ2 |
| D11 | Benefits | The benefits that stakeholders can get from using traceability in the maintenance and evolution phase | RQ3 |
| D12 | Cost | The costs of using traceability in the maintenance and evolution of software systems | RQ4 |
| D13 | Tools | The tools used for deploying the traceability practices in the maintenance and evolution phase | RQ5 |
| D14 | Approaches | The specific approaches/methods used to manage traceability in the maintenance and evolution of software systems | RQ5 |
| D15 | Challenges | Challenges of using traceability in software maintenance and evolution | RQ6 |

*3.2.5 Data Synthesis*

The data extracted in the previous step was synthesized to answer the six RQs (see TABLE IV), and the data synthesis process was conducted by following the same procedure as employed in the data extraction process. The systematic map of the extracted data is provided in Section 4.1.2.

Descriptive statistics and two coding steps of Grounded Theory (i.e., open and selective coding) were used to analyze the extracted data for answering the RQs. Specifically, descriptive analysis was adopted to answer RQ1 (maintenance and evolution activities), RQ2 (evidence level), and RQ5 (tools and approaches); open and selective coding of Grounded Theory were employed to answer RQ1 (maintenance and evolution activities), RQ3 (benefits), RQ4 (costs), and RQ6 (challenges).

We extracted the data item (D10) to answer RQ2, which collects the evidence level that traceability impacts the software maintenance and evolution activities. The evidence level can be used as an indicator of the quality and credibility of the selected studies. We adopted the following six evidence levels proposed in [44]:

Level 0. No evidence.

Level 1. Evidence obtained from demonstration or working out with toy examples.

Level 2. Evidence obtained from expert opinions or observations (e.g., survey or interview).

Level 3. Evidence obtained from academic studies (e.g., controlled lab experiments).

Level 4. Evidence obtained from industrial studies (e.g., causal case studies in an industrial setting).

Level 5. Evidence obtained from industrial practice (e.g., the solution adopted by practitioners).

Finally, the data items D11, D12, and D15 were used to extract the data for answering RQ3, RQ4, and RQ6, respectively. The extracted data was analyzed using open and selective coding of Grounded Theory, which is a well-known research method used to generate theories from qualitative data [14]. Classical Grounded Theory consists of three coding steps, i.e., open coding, selective coding, and theoretical coding [15]. We did not consider the theoretical coding step because this study was not aimed at developing a specific theory.

Open coding generates codes for incidents that can be further classified into concepts and categories. The extracted textual data was broken up and classified into various categories. In this phase, the codes for certain data items could be generated (e.g., D9: maintenance and evolution activities in TABLE IV). Selective coding identifies the core categories that explain the greatest variation in the data and around which the emerging theory is built around [14]. This phase was used to identify core categories that present the key concerns of researchers (e.g., six types of benefits that stakeholders can get using traceability in software maintenance and evolution). Moreover, constant comparison method [14] was used during the coding process to iteratively compare similarities and differences of the emerging codes and categories with the existing coding results until no more concepts and categories were generated.

## 4 STUDY RESULTS

### 4.1 Overview

In Section 4.1.1, the study search results, the selection results in each selection round, and the findings of the backward snowballing are presented. Section 4.1.2 highlights the distribution of the selected studies.

### 4.1.1 Search and Selection Results

Fig. 1 shows the number of studies in our study search, each round of study selection, and the snowballing. 31,496 papers in total were returned by the automatic search using the search terms defined in Section 3.2.1. We shortlisted 2,057 papers after the first round selection (i.e., by title). The second round search (i.e., by abstract) returned 292 studies that were further refined in the final round (i.e., by full text), where 58 primary studies were finally selected.

Moreover, two cycles of backward snowballing were conducted to explore the articles that were missed during the search process of digital databases. In each cycle, the articles were evaluated based on the title, abstract, and full text. The two cycles of backward snowballing search returned 2,366 references of the 58 studies that were finalized during the databases search. The selected 2,366 references were further assessed based on the title (shortlisted 272 articles), abstract (shortlisted 63 articles) and full text (shortlisted 5 articles) (see Fig. 1). A total of 63 primary studies were finally considered for this SMS, where 58 studies were included during the digital database search and 5 studies were finalized using the backward snowballing search approach. The details of the selected 63 primary studies are provided in Appendix A; [Sx] is used as the identifier of each primary study.

### 4.1.2 Studies Distribution

The details of the selected studies are provided in this section; the details include studies classification over publication years (D4), publication venues (D5), publication types (D6), author types (D7), maintenance and evolution activities (D9), and research types (D8).

Fig. 2 shows the yearly distribution of the selected studies published from 2000 to 2020. It should be noted that this SMS did not find any primary studies published in the first five months of 2020 (see Fig. 2) as the search was carried out during May 2020.

The details about the publication venues, types and percentage of occurrence are provided in Appendix B. The selected 63 primaries studies are scattered across 50 different publication venues. One-third of the venues (17 out of 50) are relevant to software traceability, maintenance or evolution, because their names contain the terms "traceability", "maintenance", "evolution" or the terms related to maintenance and evolution activities as described in Section 4.2.1. The results also show that all the venues published less than 5 primary studies. A majority (96.2%, 47 out of 52) of the venues published only one or two studies. These findings demonstrate that this research topic has attracted widespread concern in diverse fields in the software engineering community.

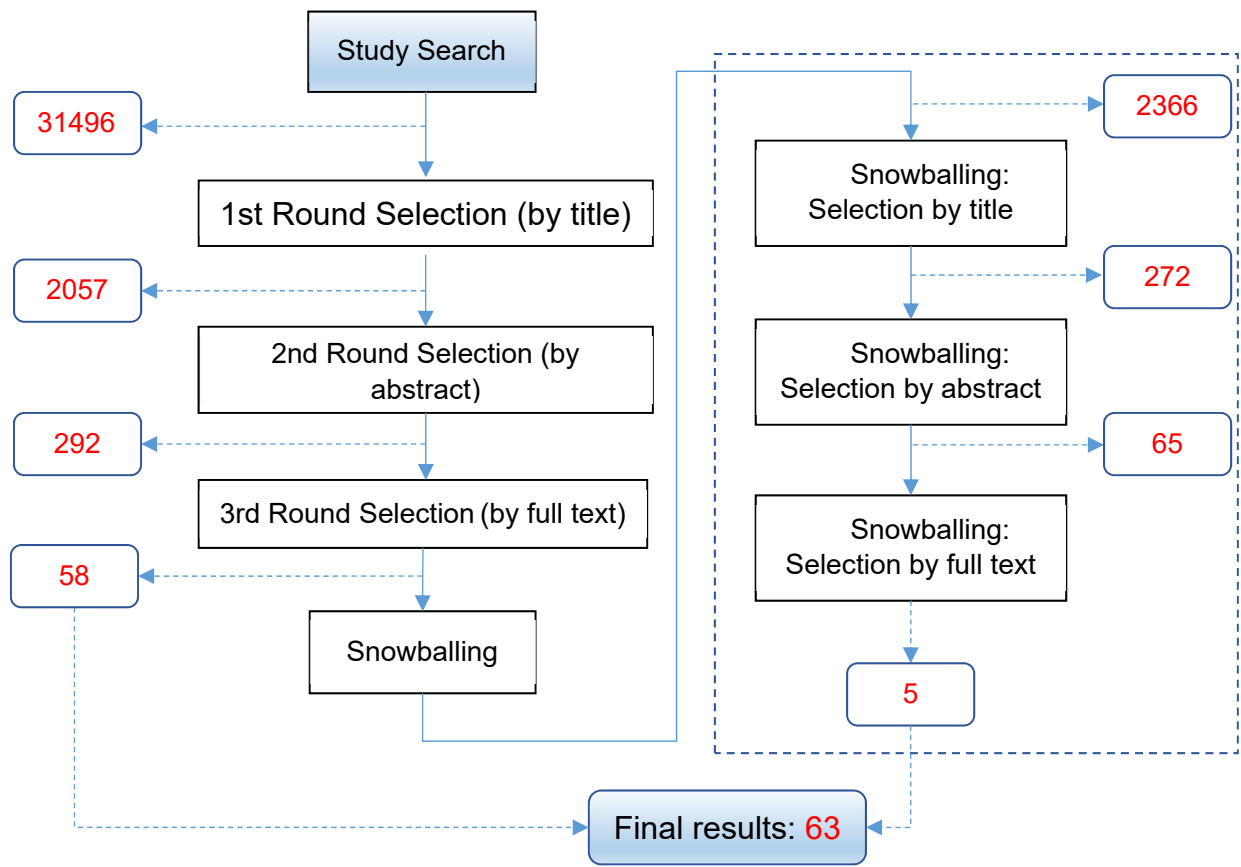

Fig. 1. Results of study search and selection.

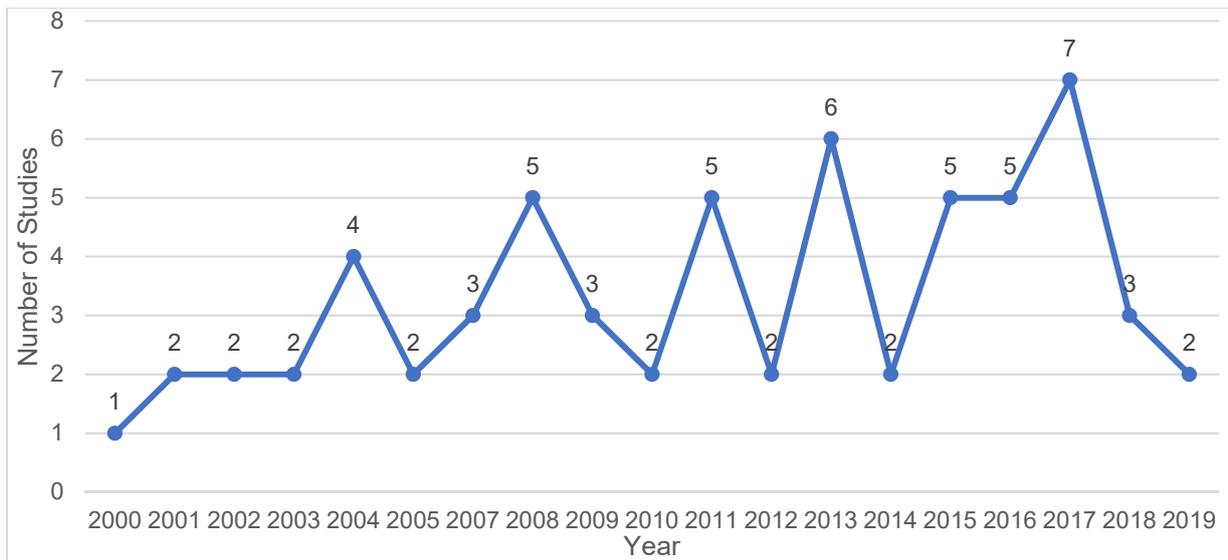

Fig. 2. Yearly distribution of the selected studies.

Furthermore, as shown in Fig. 3(a), the 50 venues cover 4 types of publications i.e., journal, conference, workshop, and book chapter. Most of the selected papers are published in journals (50.8%, 32 out 63) and conferences (28.6%, 18 out 63), however, only one paper is published as a book chapter. In addition, 19.0% (12 out of 63) primary studies are published in workshop proceedings.

Fig. 3(b) highlights the distribution of the author types (i.e., researchers, practitioners, or both) of the selected papers. Most of the papers (95%, 60 out of 63) are authored by researchers. Only 5% (3 out of 63) of the papers were based on the collaboration between researchers and practitioners. More interestingly, we do not find any paper solely authored by practitioners. The reported findings reveal that less attention has been received to the topic of traceability, software maintenance and evolution in industry as compared to academia.

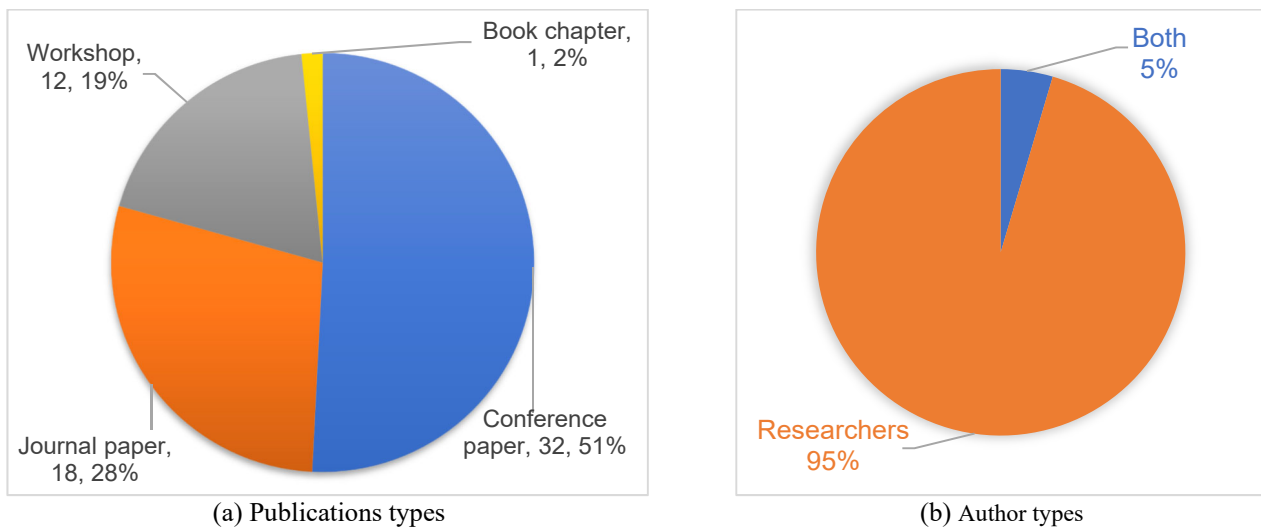

(a) Publications types    (b) Author types

Fig. 3. Publication and author types based on the distribution of the reviewed papers.

Fig. 4 presents the classification of the selected papers based on the maintenance and evolution activities (D9), research types (D8), and time period. The details of maintenance and evolution activities are reported in Section 4.2.1.

For research types, we adopted the classification scheme proposed by Wieringa *et al*. [35], who classified Requirements Engineering (RE) papers into six categories:

Evaluation research: In such type of research studies, the software engineering techniques, methods, tools or other solution is implemented or investigated in practice with research methods such as industrial case study and practitioners targeted survey.

Validation research: The properties of a proposed solution are investigated in a laboratory setting using research methods, such as mathematical analysis and laboratory experimentation.

Solution proposal: A solution technique is proposed and its relevance is discussed without a full-scale validation.

Conceptual proposal (or Philosophical paper): A new way of looking at the existing things is developed in the form of a taxonomy or conceptual framework.

Opinion paper: Personal opinion on a special matter is discussed without relying on the related work and research methodologies.

Experience report: A list of lessons learned by authors from personal experience are discussed.

It is worth noting that this classification scheme is applicable to the papers not only in Requirements Engineering but also in Software Engineering [19][36]. A decision table given in [19] is used to classify the selected secondary studies across the research types [35]. There is much confusion about the distinction between validation and evaluation research. The criterion for evaluating research papers (i.e., evaluation research or validation research) is not whether the solutions validated or evaluated are novel. The criterion is whether the solutions are implemented or evaluated in practice. For example, empirical studies that employ research methods (e.g., case study, experiment, or survey) with students are classified as validation research (e.g., [S53][S56]), while empirical studies conducted with practitioners using the same research methods are classified as evaluation research (e.g., [S50][S51]). Conceptual frameworks that are proposed without any validation and evaluation are classified as conceptual proposal (e.g., [S60]). Note that one study can span more than one research type (see Fig. 4). For example, the research reported in [S22] is considered as both opinion paper and solution proposal because the authors proposed a new solution and concluded with their opinion.

Bubbles in the left part of Fig. 4 represent the selected studies published in the context of maintenance and evolution activities with respect to the publication year. Similarly, the bubbles in the right part of Fig. 4 show the selected studies on certain research types and focusing on specific maintenance and evolution activities. The numbers shown inside the bubble represent the identification numbers of the selected studies (Appendix A). Fig. 4 shows that most of the studies are solution proposal (66.7%, 42 out of 63). Validation research is conducted in 15 (23.8%) of the selected studies; 23.8% of the papers report evaluation of solutions (i.e., techniques or approaches). Opinion papers and conceptual proposals contribute 6.3% (4 out of 63) and 4.8% (3 out of 63), respectively. We could not identify any paper as experience report study. These results indicate that only a few solutions had been used and evaluated in practice.

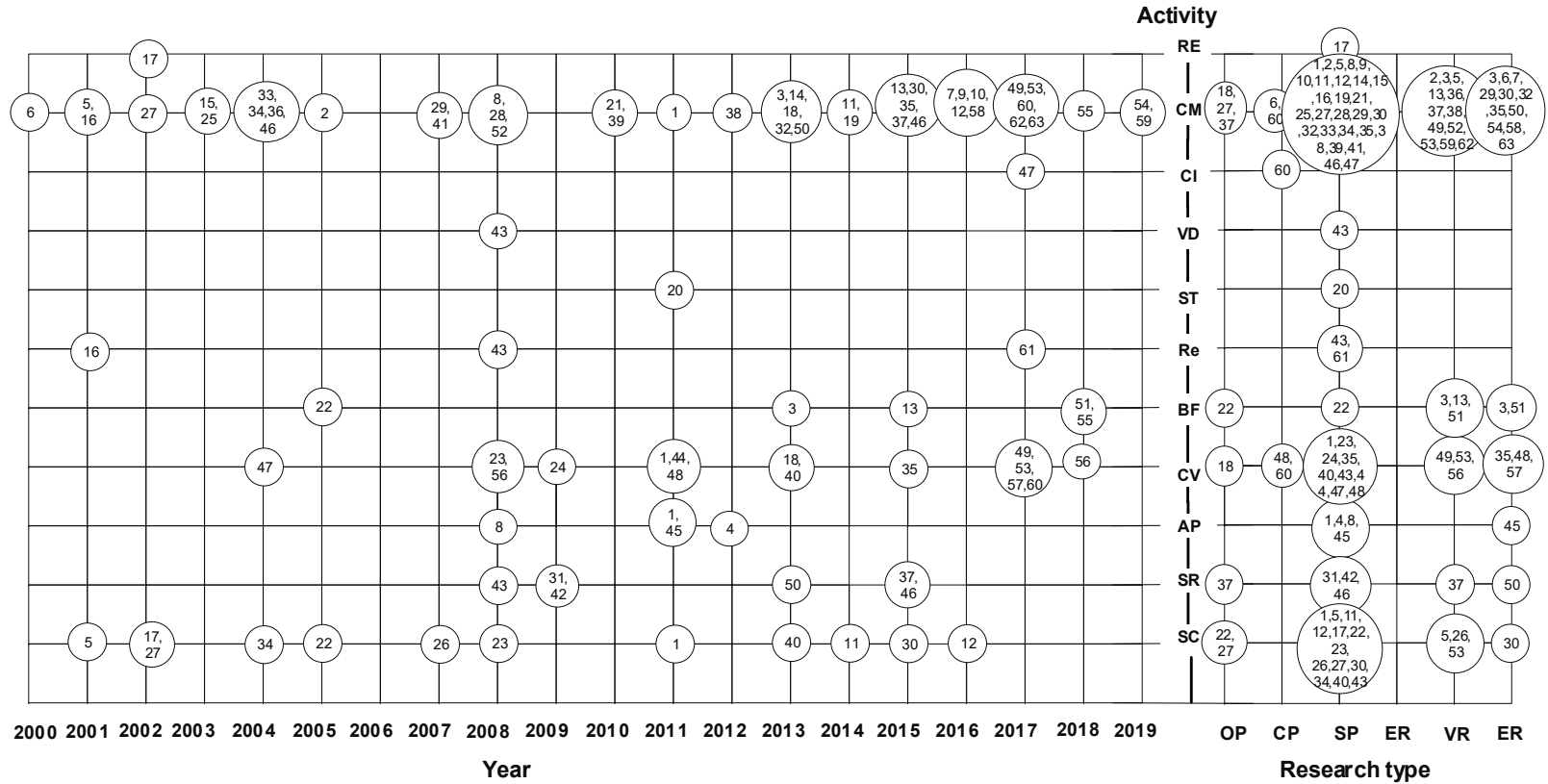

Fig. 4. Bubble chart over maintenance and evolution activities, research type, and time period.

## 4.2 Results of RQs
### 4.2.1 Maintenance and evolution activities supported by traceability (RQ1)

Lientz and Swanson categorized maintenance activities into four groups: adaptive (i.e., adapting change in software environment), perfective (i.e., updating software system with new user requirements), corrective (i.e., fixing errors), and preventive (i.e., preventing problems in the future) [1][28]. IEEE glossary considers the above-given classification as a standard for software maintenance [29]. Open coding and selective coding were used to analyze the extracted data item (D9) and identify the maintenance and evolution activities from the primary studies. For example, we generated the code "Change impact analysis" by using open coding from this sentence in [S2] "W*e would like to explore requirements traceability for change impact analysis from which we should be able to capture the impacts of a proposed change*". The generated codes were further constantly compared until no new code was identified for the maintenance and evolution activities. Finally, eleven maintenance and evolution activities that can be supported by software traceability were collected, as shown in the vertical line of Fig. 4. These activities were further classified by using selective coding across the four categories i.e., adaptive activities (change management), perfective activities (reverse engineering, software comprehension, and continuous integration), corrective activities: (bug fixes), and preventive activities (i.e., software testing, refactoring, compliance verification, software reuse, vulnerability detection, and architectural preservation). Fig. 4 further provides the details of the selected studies with respect to the 11 maintenance and evolution activities, six research types, and the time period. The frequency of occurrence of the 11 maintenance and evolution activities is shown in Fig. 5. Note that one study may contain more than one maintenance and evolution activities. Therefore, the total number of studies in Fig. 5 is larger than 63.

*Change management* plays an important role in software maintenance and evolution, since changes in software artefacts (e.g., requirements, design documents, and code dependencies) are the primary causes of software evolution [30]. The traceability of changes can support the change management process, including identifying, analyzing, evaluating, planning, implementing, and verifying the change requests [S47]. 71.4% of the primary studies (45 out of 63) mentioned that traceability can support change management. We further classified the change management activity into eight sub-activities (see TABLE V) based on the change management framework proposed in [46]. Most of the studies (84.4%, 38 out of 45) used traceability to support change impact analysis, which is a process of analyzing the consequence or ripple effects of the proposed changes [16]. For example, requirements traceability in [S12] was used to assess the impact of the prospective changes to a system. A software traceability approach [S2] was developed to support change impact analysis of object-oriented software systems. The reported approach is designed to contribute to the integration of both top-down (e.g., from requirements to low-level components) and bottom-up (e.g., from a method to its impacted test cases) impacts of system artefacts. 9.5% of the studies (6 out of 63) used traceability to identify and trace a new feature request. One study [S55] integrated requirements traceability in an Integrated Development Environment (IDE) to identify new requirements and enable their tracing to source code. Moreover, 6.3% of the studies (4 out of 63) mentioned the utilization of traceability in propagation, implementation, and regression testing of changes. For instance, a study [S21] proposed an approach and a prototype of model-based selective regression testing to support the identification of the effect of model modifications, where traceability relationships between model elements and test cases are used for test generation. The use of traceability in change effort estimation [S50], test effort estimation [S54], and change verification [S47] were analyzed in one study, respectively. Another study [S50] reports an industrial case study with the key question: *how the requirements traceability can be applied to software maintenance and evolution activities such as change effort estimation for features extensions*?

*Compliance verification* is reported as the second most common (23.8%, 15 out of 63) activity supported by traceability in maintenance and evolution. The asynchronous changes in software artefacts can lead to inconsistency between artefacts in the software evolution process. A well-defined traceability in heterogeneous artefacts is important for synchronization and consistency management among software artefacts with occurring changes in software maintenance [34]. A study [S56] presents 1.x-line mapping to support architecture-implementation traceability in a single system development and in product line development, which supports the mapping and consistency between product line architecture and code.

*Software comprehension* is considered as a significant maintenance and evolution activity [17] to swiftly understand a software system and effectively implement the requested changes. However, the total cost of software comprehension is much higher and normally it consumes half of the maintenance budget [47]. The results in Fig. 5 illustrate that 20.6% (13 out of 63) of the studies mentioned that traceability could be used to manage the software comprehension activities. Sametinger and Riebisch [S17] proposed an approach that supports evolving software systems and program comprehension by providing explicit information about dependencies and references to solution principles. Consequently, traceability can minimize the maintenance and evolution cost by reducing the program comprehension cost [S22]. Besides program comprehension, traceability could also be used to understand the meaning of program entities [S26] and the impact of the required changes on a system [S5].

*Software reuse* means adopting the existing software artefacts or knowledge, including code fragments, architectural components, patterns, and decisions to develop a new system [30]. Traceability could assist the software reuse process by minimizing the chance of selecting incorrect artefacts or missing the most suitable ones. Ultimately, it will decrease the total cost of the software maintenance process. It is also evident from the results given in Fig. 5 that 9.5% of the studies (6 out of 63) claimed traceability as a supportive tool for software reuse. Linsbauer *et al.* [S46] proposed an approach using traceability information between features and implemented artefacts to automate the reuse process of existing artefacts during the maintenance and evolution of product portfolios.

*Bug fixing* is a cost and time-consuming activity used to correct errors, flaws, or faults in a software system that produce unexpected results during the software maintenance process [48]. The result reveals that 9.5% of the studies (6 out of 63) reported traceability as an effective tool to find and fix bugs. The results of an industrial survey [S3] show that traceability links via TraceLink are effective in fixing incorrect behaviors and bugs. The results of a controlled experiment with 71 subjects in [S13] show that traceability has a significant effect on the performance of subjects when conducting maintenance tasks, including bug fixing. An approach to recover feature traceability was proposed in [S22] to reduce the effort of program comprehension and bug correction. TraceScore approach was proposed in [S51] to recover the traceability links between new bug reports, the source code, and its comments, which improves the process of defective source code files localization.

*Architectural preservation* is an activity to maintain the architectural qualities and prevent architectural erosion during the software evolution and maintenance process [49]. As shown in Fig. 5, 6.3% of the studies (4 out of 63) highlight that traceability supports the activities involved in the process of architectural preservation. Mirakhorli and Cleland-Huang [S45] presented Tactic Traceability Information Models (tTIMs) which provides a reusable infrastructure of traceability links between architectural tactics and implemented code. The use of tTIMs can assist to preserve the critical architectural qualities during the software maintenance and evolution phase.

*Refactoring* is an activity that aims at restructuring a software system to achieve the internal attributes, without changing its external behaviour [50]. In this SMS, we identified 3 primary studies that reported the use of traceability in the refactoring process. Reichmann and Müller-Glaser [S61] presented an Eclipse extension Morpheus, to develop fine granular traceability links between artefacts (e.g., source code, requirements, defects, and test cases), which supports the source code refactoring process.

Moreover, the use of traceability in *software testing, vulnerability detection, continuous integration, and reverse engineering* have been discussed in one study, respectively. For example, an Information Retrieval (IR)-based approach was used to dynamically generate multi-bidirectional links between testing documents and code to support software testing documentation [S20]. Yu *et al.* [S43] proposed a traceability technique based on a change resilience refactoring language to detect security vulnerability.

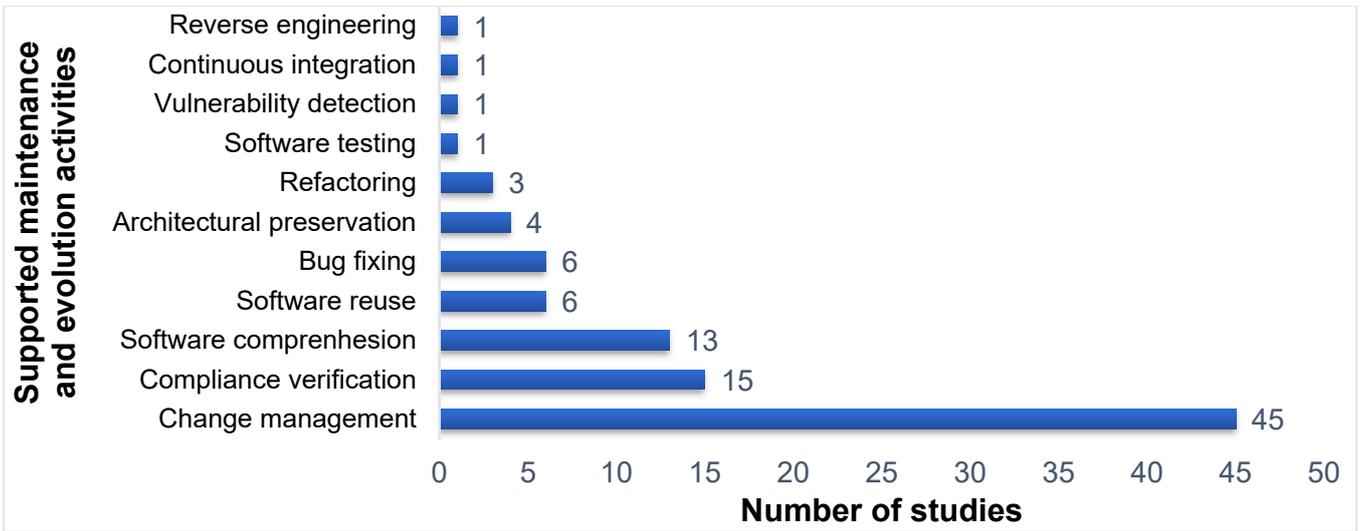

Fig. 5. Number of the reviewed studies across maintenance and evolution activities.

TABLE V. CHANGE MANAGEMENT SUB ACTIVITIES AND NUMBER OF RELATED STUDIES

| Sub activity | Studies | No. of studies | % |
|---|---|---|---|
| Change request | [S3][S13][S27][S47][S55][S63] | 6 | 9.5 |
| Change effort estimation | [S50] | 1 | 0.2 |
| Change impact analysis | [S1][S2][S5][S6][S7][S8][S9][S10][S11][S12][S14][S15][S16][S18][S19][S21][S25][S27][S28][S29][S30][S32][S34][S35][S36][S37][S38][S39][S47][S49][S50][S52][S53][S58][S59][S60][S62][S63] | 38 | 60.3 |
| Change propagation | [S33][S46][S47][S60] | 4 | 6.3 |
| Test effort estimation | [S54] | 1 | 0.2 |
| Regression testing in change | [S21][S29][S41][S54] | 4 | 6.3 |
| Change implementation | [S3][S5][S50][S58] | 4 | 6.3 |
| Change verification | [S47] | 1 | 0.2 |

*4.2.2 Evidence level concerning the impact of traceability on maintenance and evolution activities (RQ2)*

The data item D10 in TABLE IV is used to answer RQ2. The evidence level determines to what extent a specific study could be considered reliable. There are six evidence levels (as described in Section 3.2.5) and a higher evidence level means it is more likely that a study's claim is reliable.

Fig. 6 shows the distribution of selected studies across the six levels of evidence. We can see that toy example and demonstration (Level 1, 57%, 36 out of 63) is the most common approach used to evaluate traceability impact on maintenance and evolution activities. In most of these studies, the authors first introduced the details of the proposed traceability approaches or tools to support maintenance and evolution activities, and then applied these approaches and tools in some cases to show their applications. For example, [S10] proposed a traceability approach and a prototype i.e., Hybrid Coverage Analysis Tool (HYCAT). The proposed tool was further evaluated using a case study and experimentation.

Ten selected studies (16%) conducted industrial studies (e.g., industrial case studies) to validate the approaches of using traceability for maintenance and evolution activities (Level 4). Two studies ([S8] and [S25], 3%) applied the proposed traceability techniques (i.e., industrial practice) in industrial domains (Level 5). Three studies ([S18][S22][S27], 5%) reported expert insights and observations to illustrate how traceability can support maintenance and evolution activities (Level 2). It is worth noting that [S3] used both expert opinion and industrial survey to provide concrete evidence that traceability links are effective to manage the maintenance activities. In this case, we chose the higher one (i.e., industrial study) as the evidence level. Ten articles (16%) have conducted academic studies (e.g., academic controlled experiments) to evaluate the association between traceability, software maintenance and evolution process (Level 3).

Additionally, two studies ([S33] and [S46], 3%) did not provide any evidence to assess the dependencies between traceability and software maintenance (Level 0). However, these two studies mentioned without conclusive evidence that traceability is an effective tool for software maintenance and evolution.

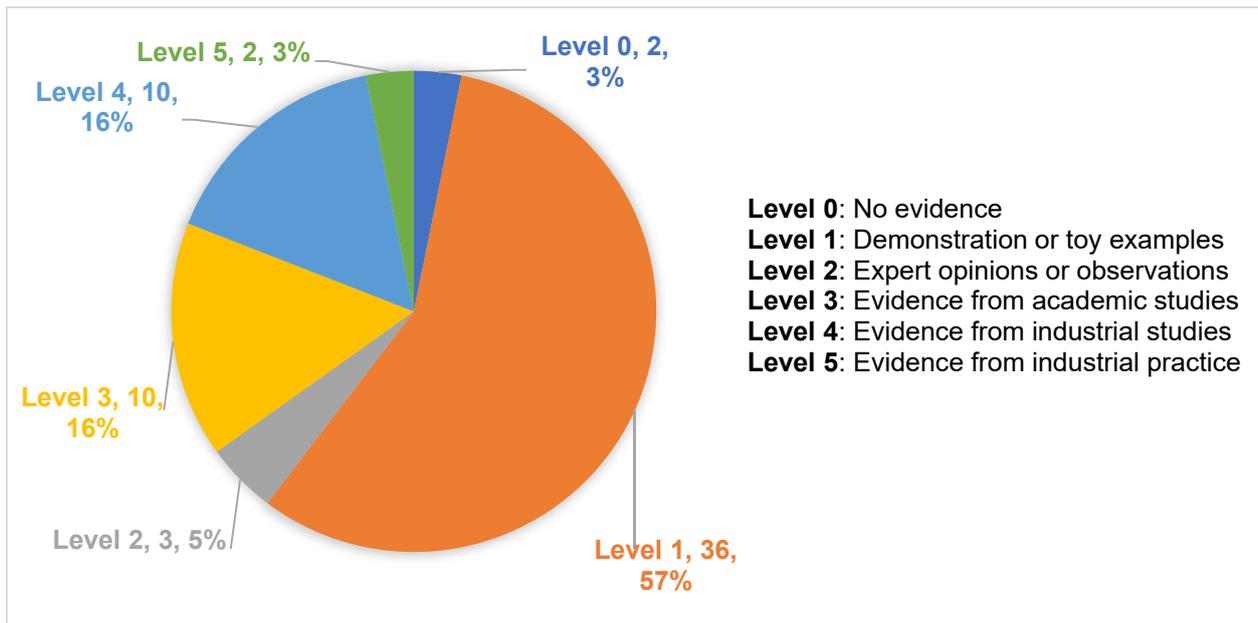

Fig. 6. Distribution of the reviewed studies across evidence levels.

4.2.3 *Benefits obtained using traceability in software maintenance and evolution (RQ3)*

The primary studies reported the benefits of using traceability for the maintenance and evolution of software systems. The identified benefits are classified into eight core categories (see TABLE VI) according to the coding steps for data synthesis as described in Section 3.2.5.

"*B1: Saving time and effort*". Software maintenance is a costly and effort consuming phase during the software development cycle. Tripathy and Naik [30] argued that 60-80% of the entire software development budget is consumed during the maintenance phase [30]. However, 15.9% (10 out of 63) of the selected studies reported that the deployment of traceability practices during the software maintenance and evolution phase could minimize the total cost, effort, and time. For example, a controlled experiment conducted by Mäder and Egyed [S13] revealed that the practitioners supported with traceability paid less time and effort to perform the maintenance and evolution activities. Their performance rate was 24% fast as compared to those who had not used the traceability practices.

"*B2: Easing the process of change management*". Traceability is considered an efficient tool for change management activities. It can improve the process of change impact analysis and assist managing the new change requests. 60.3% (38 out of 63) of the selected studies have positively described the effectiveness of traceability for analyzing the impact of the requested changes. For example, Ibrahim *et al.* [S2] proposed a requirements traceability approach that integrates requirements into low-level components, and this approach can assist developers in analyzing the impact of changes in object-oriented software systems.

"*B3: Verifying the compliance of software systems*". Traceability could be used to keep the compatibility and compliance between various artefacts during change implementation process. Tang *et al.* [S44] validated the given benefit by using traceability to co-evolve requirements specifications and architecture design.

"*B4: Preventing architecture erosion*" refers to the benefits associated with improving architecture quality and preventing architecture erosion. Mirakhorli *et al.* [S4] proposed an approach that can improve architecture attributes by deploying the traceability practices. Moreover, Javed and Zdun [S37] mentioned that the use of traceability can enhance the quality of architectural evolution analysis.

"*B5: Easing system comprehension*". Traceability plays an important role in system comprehension by visualizing and locating the dependencies between artefacts. Sametinger and Riebisch [S17] presented a supporting methodology for program comprehension by providing explicit information about the dependencies and references between patterns, aspects, and traces.

"*B6: Easing reuse of artefacts or knowledge*" refers to the benefits that can be achieved by using traceability for reuse process in maintenance and evolution. For example, the experimental results in [S37] revealed that traceability can minimize the chance of reusing irrelevant components, which enhances the architecture-centric reuse process of software systems.

"*B7: Correcting issues in software*" means that traceability can assist developers in locating the artefacts with smells, vulnerabilities, or bugs. Jurjens and Mylopoulos [S43] found that integrating traceability practices with a UML analysis tool can effectively find vulnerabilities in software systems.

"*B8: Others*" contains the benefits that are relevant to using traceability for improving software testability, reengineering process, and continuous integration.

TABLE VI. THE BENEFITS THAT STAKEHOLDERS CAN GET FROM TRACEABILITY

| ID | Benefit type | Subtype | Description | Studies |
|---|---|---|---|---|
| B1 | Saving time and effort | | The time and effort required to perform maintenance and evolution activities can be reduced using traceability practices. | [S3][S13][S14][S15][S19][S22][S32][S36][S43][S55] |
| B2 | Easing the process of change management | Improving change analysis impact | Traceability between artefacts can assist the practitioners in measuring the impact of the implemented changes. | [S1][S2][S5][S6][S7][S8][S9][S10][S11][S12][S14][S15][S16][S18][S19][S21][S25][S27][S28][S29][S30][S32][S34][S35][S36][S37][S38][S39][S47][S49][S50][S52][S53][S58][S59][S60][S62][S63] |
| | | Identifying new feature requests | Traceability used to identify the emerging changes and measure their impact on the concerned artefacts. | [S3][S13][S27][S47][S55][S63] |
| | | Facilitating change propagation | Traceability between artefacts can help developers and maintainers to facilitate the change propagation between the artefacts. | [S33][S46][S47][S60] |
| | | Implementing accurate changes | Traceability can assure the accuracy of the requested and implemented changes. | [S3][S5][S50][S58] |
| | | Easing the test of changes | Traceability between artefacts can reduce the amount of source code for test cases. | [S21][S29][S41][S54] |
| | | Others | Estimating the effort, as well as validating and verifying the implemented changes. | [S47][S50][S54] |
| B3 | Verifying the compliance of software systems | | Traceability keeps the documents up-to-date, consistent, and compatible with the relevant artefacts during the software evolution process. | [S1][S18][S23][S24][S35][S40][S43][S44][S47][S48][S49][S53][S56][S57][S60] |
| B4 | Preventing architecture erosion | Preserving architectural qualities | Traceability helps to maintain the architectural qualities during the maintenance and evolution phase. | [S1][S4][S8][S45] |
| | | Architectural consistency | Traceability helps developers to check whether an implementation is carried out correctly according to a given architectural design. | [S24][S43][S56][S57] |
| B5 | Easing system comprehension | Visualizing software artefacts | Traceability provides a project-level managerial visibility link between the system artefacts. | [S17][S22][S23][S27][S30][S34][S53] |
| | | Locating software artefacts | Traceability can help to locate the artefacts and their underlying knowledge that need to be maintained. | [S1][S5][S11][S12][S26][S40] |
| B6 | Easing reuse of artefacts or knowledge | | Traceability provides useful information about the relationships and dependencies between the system artefacts, which facilitates the reuse process of the artefacts and related knowledge. | [S31][S37][S42][S43][S46][S50] |

| B7 | Correcting issues in software systems | Locating the artefacts with bugs | Traceability can spot the artefacts containing bugs and facilitate the fixing process of the identified bugs. | [S3][S13][S22][S51][S55][S58] |
|---|---|---|---|---|
| | | Locating the artefacts with smells | Identifying the artefacts containing smells and enabling refactoring activities. | [S16][S43][S61] |
| | | Locating the artefacts with security vulnerabilities | Locating those artefacts which contain any vulnerabilities, and keeping the software system safe and secure. | [S43] |
| B8 | Others | | Traceability can improve software testability, reengineering process, and continuous integration. | [S17][S20][S60] |

*4.2.4 Costs of deploying traceability practices during software maintenance and evolution (RQ4)*

The cost types that need to be considered in the budget for implementing traceability practices were explored and extracted from the primary studies. TABLE VII shows that the identified costs are classified into four types.

"*C1: Effort for establishing and maintaining traceability links*". Employing traceability practices in the software maintenance and evolution phase is labor-intensive and time-consuming. We notice that 30.2% (22 out of 63) of the studies agree with the above statement and highlight that effort is needed to establish and maintain traceability links. For example, the ADVERT approach proposed in [S40] requires additional effort during system development for creating trace links between existing artefacts (e.g., architectural models).

"*C2: Effort for understanding traceability links*". The use of traceability involves the comprehension of traceability links between software artefacts, as well as domain knowledge for using traceability [S2][S6][S23]. For example, [S23] stated that maintainers spent effort on understanding different representations and links that exist among software artefacts and knowledge resources involved in software maintenance.

"*C3: Effort for application of traceability*". The effort is required from practitioners to integrate traceability practices with the maintenance and evolution tasks [S50][S58]. As complained by the participants in a survey [S50], too high costs are required to apply traceability in maintaining small- and medium-size projects.

"*C4: Effort for acquiring skills and experience about traceability*". An extra effort is also required to obtain the skills and knowledge for establishing and understanding traceability links as well as utilizing the traceability practices [S1][S2]. For example, [S1] mentioned that tracing architecture concerns to architecture tactics that address specific concerns requires significant effort to be an experienced architect with the domain knowledge.

TABLE VII. THE COSTS THAT STAKEHOLDERS NEED TO PAY FOR TRACEABILITY

| ID | Cost type | Description | Studies |
|---|---|---|---|
| C1 | Effort for establishing and maintaining traceability links | It is a labor-intensive, time-consuming, and error-prone process to manually establish and maintain traceability links. | [S1][S3][S4][S13][S14][S18][S20][S23][S29][S30][S33][S34][S36][S38][S40][S42][S44][S45][S51][S56][S58][S59] |
| C2 | Effort for understanding traceability links | Practitioners need to understand the interrelationships between software artefacts and the domain knowledge of using traceability for maintenance and evolution activities. | [S2][S6][S23] |
| C3 | Effort for application of traceability | Effort required to precisely develop the association between traceability, maintenance and evolution phases. | [S50][S58] |
| C4 | Effort for acquiring skills and experience about traceability | Additional skills, experience, understanding, and knowledge are acquired to utilize traceability practices correctly and effectively in maintenance and evolution activities. | [S1][S2] |

4.2.5 *Approaches and tools that support the use of traceability in maintenance and evolution phase (RQ5)*

Open coding was used to analyze the extracted data items (D13 and D14) and identify the traceability approaches and tools from the primary studies. For example, we generated the code "Information Retrieval (IR)-based approach" by using open coding from this sentence in [S50] "*The traceability method adopts an information retrieval technique for the semantic traceability and a DTDXML based technique to identify systematically all elements within and inter diagrams that are impacted by a requirement change*". The generated codes were further constantly compared until no new code was identified for the traceability approaches and tools. Finally, we identified 13 approaches and 32 tools from the selected studies that support traceability in software maintenance and evolution phase. These identified approaches and tools are listed in TABLE VIII and TABLE IX, respectively.

*a) Approaches that support the use of traceability in software maintenance and evolution*

TABLE VIII provides a list of the identified approaches, their descriptions, and the relevant maintenance and evolution activities supported by these approaches. We found that most of the reported approaches support specific maintenance and evolution activities presented in Section 4.2.1. The results show that *change management* is the most frequently supported activity by the identified traceability approaches (see TABLE VIII). Note that one study may propose more than one traceability approaches to support various maintenance and evolution activities. However, 19.0% (12 out of 63) of the selected studies do not discuss any traceability approach or technique that can be used to support the reported activities. Some of these 12 studies include industry surveys or controlled experiments, where the participants were invited to investigate the impact of traceability on maintenance and evolution activities. For instance, Mäder and Egyed [S13] conducted a controlled experiment with 71 practitioners to investigate the industrial effectiveness of requirements traceability. The results of the experiment show that traceability can save the effort and improve software maintenance quality.

Information Retrieval (IR)-based approaches are the most frequently used technique to support traceability in software maintenance and evolution (27.0%, 17 out of 63). These approaches focus on the automation of retrieving traceability links from artefacts. IR based approaches are considered the most effective technique for change management activities. For example, [S49] proposed an IR based approach for structural and semantic traceability between UML models. The proposed approach analyzed and managed the impact of changes on software requirements and design developed using UML. Additionally, IR-based approaches can be used for compliance verification, software testing, comprehension, bug fixing, and software reuse.

Feature Model (FM)-based approaches are ranked the second most commonly used technique (17.5%, 11 out of 63) to support the maintenance and evolution activities. It generally supports six activities, and change management is the most mentioned one (8 out of 11). The other activities supported by feature model-based approaches include compliance verification, software comprehension, architectural preservation, bug fixing, and software reuse. Riebisch [S34] introduced a feature model as an intermediate element to map the sets of requirements to a specific feature and structure the association between requirements, design elements, and implementation components. The industrial evaluation of the proposed approach revealed that it could improve the comprehension of change impacts made by the developers.

Scenario-based approaches are reported in eight studies that can enhance the application of traceability practices in six types of maintenance activities, and change management is the most discussed activity supported by scenario-based approaches. Two studies [S43][S44] described that the proposed approaches can also be used to facilitate compliance verification, refactoring, and bug fixing. Moreover, one study [S43] reported that the scenario-based approach can be used for software reuse and vulnerability detection, and provided a tool UMLsec to model functional and security requirements, which supports vulnerability analysis by reusing the existing test cases and integrating traceability in test cases and artefacts.

Tactic and decision-based approaches were proposed in six studies to support four maintenance activities, i.e., software comprehension (3 out of 6), architectural preservation (3 out of 6), change management (3 out of 6), and compliance verification (2 out of 6). These approaches were mainly used to understand the implemented architecture decisions or tactics, and trace changed decisions to tackle architectural problems. For example, [S4] proposed a decision-based approach using tactic Traceability Information Models (tTIMs) to connect the tactic-related classes to a relevant set of design rationales, requirements, and other related artefacts, which supports architectural preservation during the software maintenance phase.

Constraint-based approaches are regarded as efficient approaches by six primary studies, in which 3 studies specifically considered these approaches an effective technique to manage the change management activities. Moreover, constraint-based approaches are further reported as a supportive technique for software comprehension [S5], refactoring [S6], and compliance verification [S24].

Machine Learning (ML)- [S4][S26], ontology- [S23][S44], and transformation-based [S41][S56] approaches are used to support maintenance and evolution activities including software comprehension and change management. Zheng *et al.* [S56] proposed 1.x-way mapping using an architecture change model with architecture-based code regeneration to map incremental architecture changes to source code, which supports the feature-architecture-code conformance, and the case study results show that 1.x-way was applicable to maintain architecture-implementation conformance during the evolution of a real software system.

The primary studies also discussed five other approaches, i.e., goal-centric [S39], event-based ([S15], rule-based [S32], feedback-based [S12], and hypertext-based [S3] approaches. One study [S15] proposed a novel traceability approach based on the association between requirements and other artefacts using the publish-subscribe relationships for managing evolutionary changes.

TABLE VIII. THE APPROACHES THAT SUPPORT THE USE OF TRACEABILITY IN SOFTWARE MAINTENANCE AND EVOLUTION

| ID | Approach | Description | Supported activities | Studies |
|---|---|---|---|---|
| Ap1 | Information Retrieval (IR)-based approach | These approaches automatically generate traceability links from artefacts by similarity comparison between two types of artefacts with Latent Semantic Indexing, Vector Space Model, or Tribalistic Model. | Change management | [S6][S9][S14][S27][S28][S29][S35][S36][S38][S49][S59][S62] |
| | | | Compliance verification | [S35][S48][S49] |
| | | | Software comprehension | [S22][S27] |
| | | | Bug fixing | [S22][S51] |
| | | | Software testing | [S20] |
| | | | Software reuse | [S42] |
| Ap2 | Feature Model-based approach | These approaches use feature models to describe requirements and structure the variability of a product line with features as nodes and feature relations as edges. | Change management | [S8][S9][S18][S25][S34][S46][S47][S52] |
| | | | Compliance verification | [S18][S47][S56][S57] |
| | | | Software comprehension | [S22][S34] |
| | | | Architectural preservation | [S8] |
| | | | Bug fixing | [S22] |
| | | | Software reuse | [S46] |
| Ap3 | Scenario-based approach | This approach uses the hypothesized trace information to automatically generate traceability links based on observing test scenarios. | Change management | [S2][S3][S21][S33][S58] |
| | | | Compliance verification | [S43][S44] |
| | | | Refactoring | [S43][S61] |
| | | | Bug fixing | [S3][S58] |
| | | | Software reuse | [S43] |
| | | | Vulnerability detection | [S43] |
| Ap4 | Tactic and decision-based approach | These approaches provide traceability links by mapping artefacts to an implemented architectural tactic or decision. | Software comprehension | [S1][S11][S40] |
| | | | Architectural preservation | [S1][S4][S45] |
| | | | Change management | [S1][S6][S11] |
| | | | Compliance verification | [S1][S40] |
| Ap5 | Constraint-based approach | These approaches build the traceability links by using a set of constraints among different types of artefacts that must not be violated by any way. | Change management | [S5][S15][S16] |
| | | | Software comprehension | [S5] |
| | | | Refactoring | [S16] |
| | | | Compliance verification | [S24] |
| Ap6 | | | Architectural preservation | [S4] |

| | Machine Learning-based approaches | These approaches use machine learning techniques to train classifiers for detecting the tactic-related classes | Software comprehension | [S26] |
|---|---|---|---|---|
| Ap7 | Ontology-based approaches | These approaches provide formal ontological representations for both source code and document artefacts to recover traceability links between the implementation and documentation at the semantic level. | Compliance verification | [S23][S44] |
| | | | Software comprehension | [S23] |
| Ap8 | Transformation-based approaches | These approaches leverage model transformation techniques to generate artefacts (e.g., code and test) and the trace links between them. | Change management | [S41] |
| | | | Compliance verification | [S56] |
| Ap9 | Goal-centric approaches | These approaches allow the stakeholders to evaluate the impact of changing functional requirements on non-functional requirements. | Change management | [S39] |
| Ap10 | Event-based approaches | These approaches use publisher-subscriber relationships between the artefacts based on the event notifications to update and maintain traceability relationships. | Change management | [S15] |
| Ap11 | Rule-based approaches | These approaches can automatically generate traceability links using rules based on the attributes of the artefacts. | Change management | [S32] |
| Ap12 | Feedback-based approaches | These approaches generate traceability links by integrating feedback into the implemented requirements to trace the system evolution. | Change management Software comprehension | [S12] |
| Ap13 | Hypertext-based approaches | These approaches mainly create the traceability between requirements and source code using XML. | Feature extension Change management Bug fixing | [S3] |

*b) Tools used to support the traceability practices in software maintenance and evolution*

Tools are essential to (semi-)automatically establish, discover, and maintain traceability links during the evolution of software systems. In this SMS, we explored the primary studies and collected the identified traceability tools. The details of the tools and the list of the supported activities are provided in TABLE IX. Note that one tool can be used to support more than one traceability approach and maintenance and evolution activities. Moreover, some studies may not provide any details about the tools (e.g., [S54]), on the other hand, some studies may report more than one tool (e.g., [S2]). We notice that 44.4% of the studies did not discuss any specific tools for supporting traceability.

55.6% (35 out of 63) of the studies reported 32 tools that support the use of traceability in maintenance and evolution activities. Except for four tools (i.e., Text editor, D3TraceView, HYCAT, SAT-Analyzer) that are not explicitly mentioned in the studies to support traceability approaches, the rest of the identified tools support at least one traceability approach and maintenance and evolution activity. We notice that 6 tools support three maintenance and evolution activities and 13 tools facilitate two activities. These results show that tools have been widely used to bridge the gap between traceability and software maintenance and evolution. The results in Fig. 7 illustrate that ten maintenance and evolution activities are supported by the identified tools: 60.0% (21 out of 35) of the tools are used to support change management activity; compliance verification activity is supported by 37.1% (13 out of 35) of the tools; and software comprehension receives 20.0% (7 out of 35) tools support.

There are no tools for supporting the use of traceability in software testing activity. Moreover, the results in Fig. 8 demonstrate that 11 traceability approaches can be supported by using the identified tools. IR-based and scenario-based approaches are the most common techniques that are supported by 25.7% (9 out of 35) and 20% (7 out of 35) of the tools, respectively.

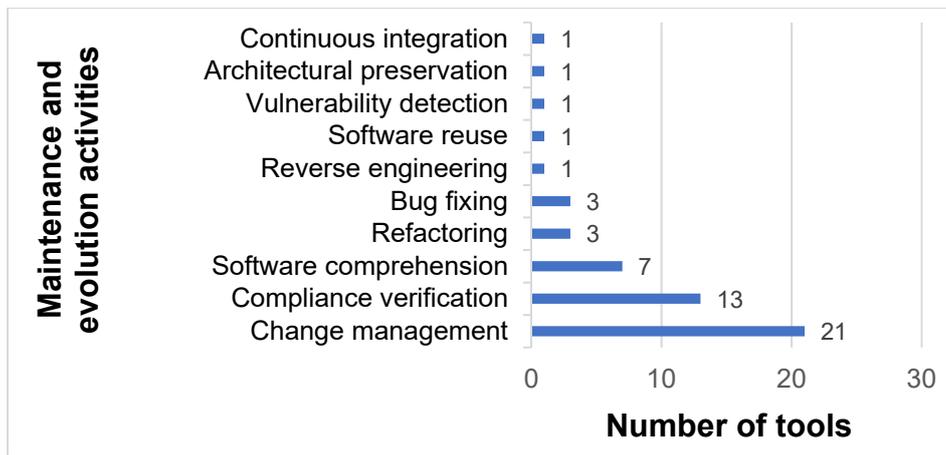

Fig. 7. Number of tools supporting maintenance and evolution activities.

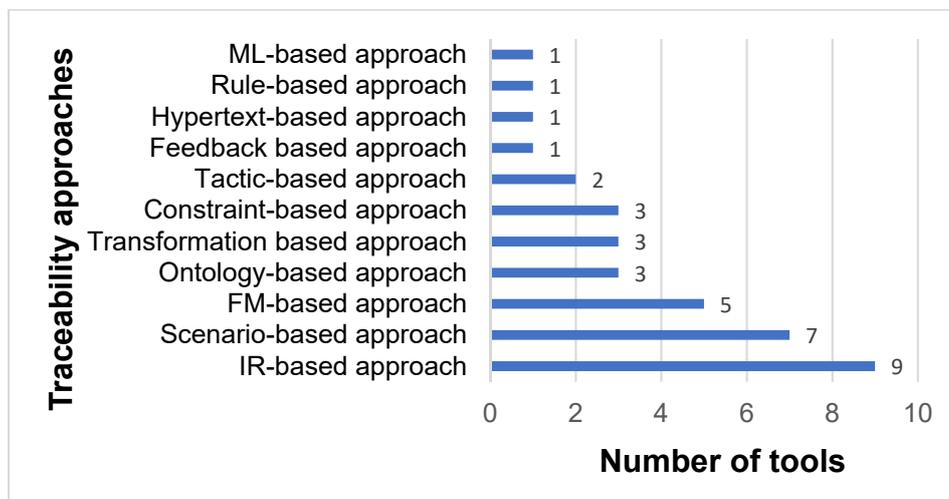

Fig. 8. Number of tools supporting traceability approaches.

TABLE IX. THE TOOLS THAT SUPPORT TRACEABILITY USED IN SOFTWARE MAINTENANCE AND EVOLUTION

| ID | Tool | Studies | Traceability approach | Maintenance and evolution activities |
|---|---|---|---|---|
| T1 | Eclipse plugin | [S26][S39][S41][S55] | ML-based approach; Transformation-based approach | Bug fixing; Change management; Software comprehension |
| T2 | CQV-UML | [S49][S59][S62] | IR-based approach | Change management; Compliance verification |
| T3 | IR-based tool | [S14][S35] | IR-based approach | Change management; Compliance verification |
| T4 | CASE tool | [S16][S48] | Constraint-based approach; IR-based approach | Change management; Compliance verification; Refactoring |
| T5 | xLineMapper | [S56][S57] | Transformation-based approach; FM-based approach | Compliance verification |
| T6 | Code parser | [S2][S23] | Scenario-based approach; Ontology-based approach | Change management; Compliance verification; Software comprehension |
| T7 | Ontology editor | [S23] | Ontology-based approach | Compliance verification, Software comprehension |
| T8 | MbSRT$^2$ | [S21] | Scenario-based approach | Change management |
| T9 | Text editor | [S13] | | Bug fixing; Change management |
| T10 | Morpheus | [S61] | Scenario-based approach | Refactoring |
| T11 | tTIM | [S45] | Tactic-based approach | Architectural preservation |
| T12 | SE-Wiki | [S44] | Ontology-based approach; Scenario-based approach | Compliance verification |
| T13 | CMU tool suite | [S27] | IR-based approach | Change management; Software comprehension |
| T14 | Archface compiler | [S24] | Constraint-based approach | Compliance verification |
| T15 | Integrate | [S19] | FM-based approach | Change management |
| T16 | ANALYST | [S6] | Tactic-based approach; IR-based approach | Change management |
| T17 | Catia | [S2] | Scenario-based approach | Change management |
| T18 | xMapper | [S56] | FM-based approach; Transformation-based approach | Compliance verification |
| T19 | D3TraceView | [S53] | | Change management; Compliance verification; Software comprehension |
| T20 | ART | [S43] | Scenario-based approach | Vulnerability detection; Refactoring; Compliance verification; Software reuse |
| T21 | TIRT | [S38] | IR-based approach | Change management |
| T22 | RETRO | [S29] | IR-based approach | Change management |
| T23 | QuaTrace | [S25] | FM-based approach | Change management |
| T24 | Javadoc extension | [S17] | Constraint-based approach | Reverse engineering; Software comprehension |
| T25 | CAFÉ | [S12] | Feedback based approach | Software comprehension |
| T26 | HYCAT | [S10] | | Change management |
| T27 | Evo-SPL | [S9] | IR-based approach; FM-based approach | Change management |
| T28 | TraceLink | [S3] | Hypertext-based approach | Change management; Bug fixing |
| T29 | EMFTrace | [S32] | Rule-based approach | Change management |
| T30 | SAT-Analyzer | [S60] | | Change management; Compliance verification; Continuous integration |
| T31 | MRTA | [S48] | IR-based approach | Compliance verification |
| T32 | CodeMentor | [S2] | Scenario-based approach | Change management |

### 4.2.6 Challenges of using traceability practices in software maintenance and evolution (RQ6)

The reviewed studies clarified that there can be various challenges in implementing traceability practices in maintenance and evolution as shown in TABLE X, which are detailed below.

"*CH1: Quality of traceability links*" is reported as a challenge in 22% (14 out of 63) of the studies. It is challenging to assure the quality of traceability practices because of certain factors, including unclear traceability, lack of traceability knowledge, and coarse granularity of traceability. For example, Charalampidou *et al.* [S55] argued that it was quite challenging to successfully deploy the traceability practices between system artefacts and develop a structured roadmap to link requirements to test cases, which makes it difficult to facilitate certain activities (e.g., fixing bug, adding features, and locating artefacts).

"*CH2: Performance of traceability approaches and tools*": The results highlight that 22% (14 out of 63) of the studies considered the performance assurance of the traceability tools and techniques as a serious challenge. The poor performance of the tools and techniques could be a disaster for implementing the traceability process. One study [S35] adopted an information retrieval approach to trace and update the requirements that are impacted during the change implementation process. However, this approach missed some relevant and impacted requirements, which leads to inconsistency in requirements specification.

"*CH3: Creation of traceability links*" denotes the difficulties of establishing traceability links that support specific software maintenance and evolution activities. These challenges include how to automatically create trace links, extract the implicit links or dynamic links, and create links between system artefacts. One study [S56] proposed the 1.x-way mapping that semi-automatically built traceability from architecture changes to source code for maintaining the architecture-implementation conformance in both single system and product line development. However, the 1.x-way mapping cannot model or automatically enforce the relationships (e.g., mutual dependency) between product line features.

"*CH4: Predicting traceability cost and benefits*": It is hard to predict or evaluate the effort and cost of using traceability during the maintenance and evolution phase [31]. We present in Section 4.2.3 and Section 4.2.4 respectively regarding the benefits and costs of using traceability in maintenance and evolution. One study [S6] proposed that a fine-grained level of traceability in software representation model improved the effectiveness of impact analysis, but also increased the effort of maintenance. This study concluded that a cost-benefit analysis is needed to measure the effectiveness of impact analysis.

"*CH5: Lack of tool support*": We list various tools in Section 4.2.5 that can support the use of traceability in software maintenance and evolution. However, it is still difficult to employ these tools in real-world environment because the traceability information needs to be exchanged and integrated across software projects [31]. It is evident from the results that only four studies ([S26][S39][S41][S55]) reported traceability tools as Eclipse plugins to support the use of traceability in maintenance and evolution activities. As mentioned in [S55], all the participants agreed that external traceability tools are difficult to be integrated into the development process and environments of developers (e.g., Git and Eclipse).

TABLE X. THE CHALLENGES OF USING TRACEABILITY IN SOFTWARE MAINTENANCE AND EVOLUTION

| ID | Challenge | Description | Studies |
| --- | --- | --- | --- |
| CH1 | Quality of traceability links | The performance of using traceability is mainly based on the quality of traceability links, which is critical to use traceability in maintenance and evolution activities. | [S7][S15][S22][S24][S25][S35][S36][S41][S42][S51][S55][S59][S62][S63] |
| CH2 | Performance of using traceability approaches and tools | Inefficient use of traceability tools and techniques could negatively impact the benefits and costs of employing traceability in maintenance and evolution. | [S2][S4][S9][S23][S27][S28][S35][S36][S42][S45][S48][S49][S59][S62] |
| CH3 | Creation of traceability links | It is challenging to build traceability between a large number of software artefacts involved in the maintenance and evolution phase. | [S7][S14][S18][S24][S40][S51][S55][S56][S57][S59] |
| CH4 | Predicting the costs and benefits of using traceability | It is hard to predict the effort and Return on Investment (ROI) of using traceability, and there is a need of dedicated metrics for measuring the benefits, cost, and effort. | [S6][S7][S13][S45][S50] |

| CH5 | Lack of tool support | There is a lack of dedicated tools to use and integrate traceability approaches in specific maintenance and evolution activities in the development environment. | [S7][S18][S55] |

## 5 DISCUSSION

### 5.1 Analysis of Study Results

**Maintenance and evolution activities**: We identified 11 maintenance and evolution activities that can be supported by using traceability (see Section 4.2.1). More specifically, 71.4% (45 out of 63) of the reviewed studies indicate that traceability can support *change management* activity in the maintenance and evolution phase. The selected studies mostly used traceability during change impact analysis. However, the result is not surprising as changes are always expected from the requirements gathering to the maintenance phase of the software development life cycle. Moreover, changes impact source code and other related artefacts e.g., requirements, design, and test cases, where traceability plays a key role to analyze and understand the significance of the changes [43].

*Compliance verification* is ranked as the second most common (23.8%, 15 out of 63) activity of the maintenance and evolution phase that can be improved using the traceability practices. The result further confirms that consistency between software artefacts can be enhanced using traceability. Moreover, *software comprehension* is the third most frequently reported maintenance and evolution activity supported by traceability (20.6%, 13 out of 63). The evolution of complex software systems requires an understanding of the changes and dependencies between key artefacts [24]. Traceability provides the dependencies between various artefacts that play a critical role in system understanding, e.g., program comprehension. Using traceability, developers can easily trace code to the original requirements which provides the rationale for implementation and trace code back to design and architecture which gets a high-level view of a system [42]. Furthermore, very few studies used traceability to support software testing, vulnerability detection, continuous integration, and reverse engineering. For example, [S61] highlights the lack of traceability management techniques in continuous integration due to the limitation that the current traceability approaches mainly address the requirements and design level artefacts without covering the artefacts in the later phases of software development life cycle (e.g., test reports and configuration files).

Based on the results and analysis of the reviewed studies, we can conclude that *traceability practices have a comprehensive and positive impact on software maintenance and evolution as 11 maintenance and evolution activities can be supported by using traceability practices*.

**Evidence level**: The results presented in Section 4.2.2 reveal that the majority (81.0%, 51 out of 63) of the studies have not evaluated the deployment of traceability practices in industrial settings (i.e., Level 4 and Level 5). We noticed that 73.0% (46 out 63) of the studies used toy examples (Level 1) and lab experiments (Level 3) to validate the significance of traceability in the maintenance and evolution phase. Relatively less attention has been paid to using expert opinions or observation (Level 2) for evaluating the impact of traceability. Moreover, 12 studies (19.0%) provided the evidence from the industrial practices, where, only 2 studies reported the adoption of traceability and evaluated the effectiveness of traceability in industrial practices (Level 5), and the other 10 studies evaluated the traceability approaches with industrial studies (Level 4). One potential reason might be that it is difficult to measure the ROI for adopting traceability in daily maintenance activities [31]. The results regarding the evidence level are consistent with the results of the research type (see Fig. 4) as both results reveal that most of the studies only proposed traceability solutions without adopting or evaluating them in the industrial practices and studies. These results show that *the strength of the evidence on the impact of traceability in maintenance and evolution is still not strong enough*.

**Benefits and costs**: We identified and reported the key benefits of using traceability for maintenance and evolution activities. The identified benefits are further classified across eight categories (see TABLE VI). The most frequently reported benefit of using traceability is saving time and effort when conducting the maintenance and evolution tasks. The other major benefit is easing the change management process. However, it is also required to allocate specific budget to employ the traceability practices. In this SMS, we identified four types of costs that are mandatory for deploying the traceability practices (see TABLE VII). The major cost involves establishing and maintaining the traceability links (34.9%, 22 out of 63 studies). The other three cost types are the effort for understanding of traceability links and application of traceability as well as acquiring required skills and knowledge about traceability. The results show that ***the majority of the costs is related to establishing and maintaining traceability links***, and one potential reason is that establishing and maintaining traceability links is the prerequisite for understanding and applying the traceability links in maintenance and evolution activities.

**Approaches and tools**: We identified 13 approaches and 32 tools that can be used to support the traceability practices (see TABLE VIII and TABLE IX). Most of the approaches can be used to support more than one specific maintenance and evolution activities. Change management is considered the most common activity supported by the traceability approaches and tools. Moreover, most of the identified tools can support at least one traceability approach and one specific maintenance and evolution activity. However, we notice that a majority of the approaches and tools were not frequently used in industrial settings (see the evidence levels of the studies in Section 4.2.2). Due to the lack of industrial evidence, the reported approaches and tools might not be effective when used in practice (for example, one of the identified challenges is *CH2: Performance of traceability approaches and tools* as presented in Section 4.2.6). ***The results clearly highlight the gap between industry and academia regarding the available approaches and tools that support traceability in software maintenance and evolution***.

**Challenges**: The selected studies were explored to identify the key challenges that can be the potential barriers to use traceability for maintenance and evolution activities. We identified five types of challenges (see TABLE X), in which ***three main challenges are related to the development of traceability links and the quality of and performance of using the links***. Manual creation of trace links is costly; however automated trace recovery induces incorrect trace links. Automatically ensuring accurate and change-resilient traceability [S43] is difficult because trace links have to be rediscovered when changes happen in related artefacts during software evolution. Moreover, it is still difficult to exactly determine the traceability ROI figures. The effectiveness of maintenance and evolution activities can be impacted by varying granularity of traceability models [S6]. The greater effort devoted to a finer grained model, the better accuracy of change impact analysis can be achieved. A cost-benefit analysis should be accomplished based on maintenance and evolution effectiveness and the effort required for deploying traceability practices.

## 5.2 *Implications for Researchers and Practitioners*

1) The results of RQ1 show that, although using traceability can support 11 maintenance and evolution activities, there is a lack of studies that discuss how using traceability can support software testing, vulnerability detection, continuous integration, and reverse engineering. Therefore, researchers can propose dedicated approaches or tools to support using traceability in these maintenance and evolution activities.

2) The results of RQ2 show that most of the studies have provided the evidence using toy examples or academic studies. This finding indicates that the proposed solutions of using traceability in maintenance and evolution phase have been rarely evaluated in industrial settings, which indicates that there is a need of improving the real-world application and evaluation of traceability approaches and tools. For example, an industrial survey with the practitioners can be conducted to investigate the pros and cons of adopting traceability in maintenance and evolution. Therefore, industrial controlled experiments can be conducted to assess the effectiveness of the proposed traceability approaches in maintenance and evolution. Strong industrial evidence on this topic will motivate practitioners to employ traceability practices in maintenance and evolution activities.

3) It is important to analyze the costs and benefits, which is a critical part of the impact, of using traceability in maintenance and evolution phase. The results of RQ3 show that the investment required to develop more effective traceability tools could be balanced with ROI, and the investment partially depends on the accuracy of the traceability links, i.e., a higher accuracy will need more cost and effort. It is also evident from the results that the balance between the cost and benefits of using traceability in various maintenance activities requires further industrial investigation. Therefore, the future studies shall propose methods to quantitatively measure the cost-benefit ratio of using traceability in these maintenance and evolution activities. Practitioners and researchers are also encouraged to use empirical methods to evaluate the cost-benefit ratio of adopting the traceability practices in maintenance and evolution.

4) The results of RQ4 show that traceability costs mainly involve employing the traceability practices in maintenance and evolution. Most of the organizations still hesitate to consider traceability in development because of the cost factor [S48]. Therefore, researchers can utilize the existing techniques from, e.g., machine learning and natural language processing, to automate the traceability practices and minimize the total cost and effort.

5) The results of RQ5 show that 13 of the identified approaches can support the deployment of traceability practices in software maintenance and evolution phase. Most of the proposed approaches can be used to facilitate more than one maintenance and evolution activities. Therefore, we suggest that practitioners can choose suitable traceability approaches based on the activities to be supported as well as the aims and conditions of the projects. For example, software testing can be supported by Information Retrieval (IR)-based approach (see TABLE VIII), in that multi-bidirectional links between testing documents and code can be dynamically generated by using Latent Semantic Indexing (LSI) [S20].

6) The results of RQ5 show that 32 tools were identified that can be used to support traceability in maintenance and evolution activities. However, there is still lack of tools that can be used for certain traceability approaches (e.g., goal-centric and event-based approach) and maintenance and evolution activities (e.g., software testing and vulnerability detection). Therefore, we encourage researchers and practitioners to collaborate on developing such tools which can be used in practice. Furthermore, industrial studies are required to validate the effectiveness of these tools, which can help researcher and practitioners to select the most suitable tools.

7) The identified traceability challenges provide a check-list that should be considered before using traceability in maintenance and evolution. The results of RQ6 show that more effective traceability approaches and tools should be proposed to ensure the quality of traceability links for maintenance and evolution activities. Accurate and change-resilient traceability [S43] is required to be integrated with maintenance and evolution activities (e.g., software testing and vulnerability detection). Therefore, there is a need of cost-benefit metrics to analyze the ROI of using traceability in maintenance and evolution activities. Furthermore, value-oriented or activity-oriented traceability approaches and tools can be developed to support specific maintenance and evolution activities.

6 THREATS TO VALIDITY

The threats to validity of this SMS are discussed by following the guidelines provided by Shull *et al.* [25] as well as the measures taken to alleviate the identified threats.

**Construct validity** focuses on whether the theoretical constructs are interpreted and measured correctly. In this SMS, one potential threat to the construct validity is the correctness of the included primary studies. Inappropriate search terms and strategies may result in retrieving a large number of irrelevant studies and missing potentially relevant articles. To alleviate this threat, we conducted a pilot search to improve the appropriateness of the search terms. Another threat to construct validity is the interpersonal biases between the authors to decide whether a study should be retained or not during the study selection process. To alleviate this threat, a pilot selection process was firstly conducted to help all the authors to understand the selection criteria and manage the disagreements factors. Consensus was developed based on the pilot search results. Finally, the first and second authors conducted the formal studies selection process and the debatable papers were further discussed by all the authors to make the final decision about the search results.

**Internal validity** focuses on the study design and particularly whether the results match the collected data. One potential threat to the internal validity of this study is the quality of the data extracted to answer the RQs. This threat was partially alleviated by conducting pilot data extraction with ten primary studies. The pilot data extraction results were discussed by all the authors to achieve an agreement. The formal data extraction and analysis were conducted by the first and second authors; the third and fourth authors reviewed the extracted data and the analyzed results. Disagreements were discussed and resolved among all the authors. In addition, we adopted both descriptive and qualitative analysis methods (i.e., two coding steps of grounded theory) to analyze the results and mitigate this threat.

**External validity** refers to the degree at which the findings of a study can be generalized. The results of this SMS provide an overview of the existing literature on using traceability in software maintenance and evolution. To ensure the completeness of the selected studies, we conducted the search process by including seven most popular digital databases that publish software engineering related studies [8]. Moreover, the snowballing technique was employed in study search to mitigate the possibility of missing relevant studies.

**Reliability** refers to whether the findings of a study are reliable when replicated by other researchers. In this SMS, the search, selection, and data extraction and analysis were manually conducted by following the designed protocol. A potential threat to the reliability of the SMS is the interpersonal bias in data extraction and analysis. To alleviate this threat, we made the data extraction and analysis process explicit (see Section 3.2.4 and Section 3.2.5), and also took measures (e.g., pilot data extraction and analysis) to mitigate the personal bias between the authors.

## 7 RELATED SECONDARY STUDIES

Several secondary studies (i.e., SMSs and SLRs) are published concerning software traceability. Javed and Zdun [20] conducted an SMS of the existing traceability approaches between software architecture and source code. They finally selected 11 primary studies for further analysis and discussions. A classification scheme was developed to distinguish various aspects of traceability between software architecture and source code. Their findings revealed that there is a need of strong empirical evidence for software architecture traceability approaches, and proper support with widely accepted tools for completeness and consistency checking as well as querying mechanisms and standards.

Duarte *et al.* [32] proposed TraceBoK, a body of knowledge on requirements traceability by systematically reviewing 26 relevant studies to synthesize the existing approaches in requirements traceability. They further conducted an industrial survey with the experts to empirically evaluate the identified traceability approaches in TraceBoK. The survey results confirmed that TraceBoK met practitioners' needs for managing requirements traceability.

Nair *et al.* [33] conducted a review of traceability in the requirements engineering phase by including 70 papers published in the Requirements Engineering (RE) conference in the last 20 years, and the results show that lack of knowledge and understanding about traceability is the most frequently quoted challenge. Both reviews [32][33] focused on the primary studies on requirements traceability.

Borg *et al.* [21] conducted an SMS with 79 papers focusing on Information Retrieval (IR)-based trace recovery approaches and tools. They investigated the empirical evidences of the IR-based trace recovery with 132 empirical evaluations, e.g., experiments, case studies, and concluded that it is critical to improve the overall quality of the identified parameters of the IR-based trace recovery by conducting industrial case studies.

Vale *et al.* [22] conducted an SMS on traceability in Software Product Lines (SPL) for synthesizing the available evidence, research challenges, and open issues for further research. The study provides a structured understanding of SPL traceability. The results indicate that there is a lack of evidence regarding the application of traceability methods in practice, and the complex nature of variability in SPL is the main challenge for SPL traceability activities.

Similar to our work, Charalampidou *et al.* [45] conducted an SMS on 155 primary studies to analyse traceability approaches in software development in four aspects, i.e., the types of artefacts connected with traceability approaches, the goals of using traceability approaches, the quality attributes derived from using traceability approaches, and the research methods used for validating traceability approaches. Their findings highlight that the use of traceability approaches could mainly improve maintenance-related quality attributes, i.e., modifiability, correctness, instability, and understandability, while our work presents the impact of traceability on maintenance and evolution from various aspects, including the activities, empirical evidence, traceability approaches and tools, benefits, costs, and challenges.

An overview of the differences between the aforementioned secondary studies and our work is presented in TABLE XI. Several key aspects are differentiated as follows:

**Study objectives**: The SMS [20] identified the existing traceability approaches and tools between architecture and source code, as well as the empirical evidence, benefits, and liability of these approaches. The SLR [32] evaluated the traceability approaches on software product traceability. The SLR [33] presented the traceability research focuses, maturity, challenges, and topic evolution at the RE conference. The SMS [21] analyzed Information Retrieval (IR)-based traceability recovery approaches. The SMS [45] explored the goals of existing software traceability approaches and their empirical evaluation methods. Different from these secondary studies, our SMS intends to analyze the impact of software traceability on software maintenance and evolution activities.

**Traceability types**: The traceability types investigated in our SMS cover the traceability in software development, including architecture traceability [20], software product traceability [32], and requirements traceability [33]. However, two SMSs [21][45] also focused on software traceability, in which one SMS [21] exclusively explored IR-based approaches to software traceability and the other SMS [45] analyzed software traceability in empirical studies. In contrast, our SMS focuses on software traceability that impacts software maintenance and evolution activities.

**Research questions**: The RQs of our SMS are different from the RQs investigated in other secondary studies. RQ1, RQ3, and RQ4 aim to explore respectively software maintenance and evolution activities, the benefits, and costs of using traceability in maintenance and evolution activities. To the best of our knowledge, no secondary studies addressed similar RQs as RQ1, RQ3, and RQ4. Although several studies investigated similar RQs related to RQ2 (empirical evidence [21][33][45]), RQ5 (approaches [20][32] and tools [20][33]), and RQ6 (challenges [33]). Our SMS specially explores the primary studies about software traceability in maintenance and evolution activities to answer these RQs.

**Study results**: The results of our RQs are mostly different from the results of the related secondary studies (see TABLE XI). For example, our results present eleven software maintenance and evolution activities that can be supported by using traceability (RQ1) as well as eight types of benefits (RQ3) and four types of costs (RQ4) of using traceability during software maintenance and evolution phase. The results of the SMS [20] also reported that traceability approaches can benefit developers and maintainers for identifying change impacts and understanding the systems. These two types of benefits are a bit overlapped with two types of the identified benefits (i.e., B3 and B5 in TABLE VI) in our SMS. Moreover, several traceability approaches (see TABLE VIII) collected in our SMS are also reported in the SMS [20] on architecture traceability, the SLR [32] on software product traceability, and the SMS [21] on IR-based traceability, e.g., event-based approaches, rule-based approaches, goal-centric approaches, and IR-based approaches. None of the traceability tools reported in our SMS (see TABLE IX) have been presented as the traceability tools between software architecture and source code in the SMS [20] on architecture traceability. Several challenges (see TABLE X) identified in our SMS are also reported in other secondary studies. For example, CH1 is similar to the challenge of traceability knowledge understanding and traceability information representation presented in [33]; CH4 is similar to the challenge of reducing additional cost collected in [20][33]; and CH5 is similar to the challenge of needing advance tool support in industry identified in [33]. Similar to our results, the secondary studies [20][33] also suggested employing more advanced empirical methods (e.g., case study and survey) for evaluating software traceability approaches and tools. Although there are some overlaps with the related secondary studies in terms of the benefits, traceability approaches, and challenges, our SMS reports these aspects in the context of using software traceability in maintenance and evolution activities. For example, we present the mapping of traceability approaches and tools to various types of maintenance and evolution activities (see TABLE VIII and TABLE IX).

TABLE XI. COMPARISON OF RELATED SECONDARY STUDIES WITH OUR SMS

| Existing secondary studies | Traceability type | Time period | No. of studies | Queried databases | Objective |
|---|---|---|---|---|---|

| Javed and Zdun [20] | Traceability between architecture and source code | 1999 to 2013 | 11 | ACM Digital Library | Identifying the traceability approaches and tools between architecture and source code |
| --- | --- | --- | --- | --- | --- |
| Duarte et al. [32] | Software product traceability | 2007 to 2014 | 26 | IEEE Xplore, ACM Digital Library, Spring Link, Science Direct | Evaluating the traceability approaches focusing on software product traceability |
| Nair et al. [33] | Requirements traceability | 1993 to 2012 | 70 | IEEE Xplore | Evaluating how traceability research published at the RE conference has contributed to the requirements engineering area. |
| Borg et al. [21] | Software traceability | 1999 to 2011 | 79 | IEEE Xplore, Web of Science, EI Compendex, ACM Digital Library, SciVerse Hub Beta, Google Scholar | Analyzing Information Retrieval (IR)-based traceability recovery approaches with a focus on evaluation and strength of evidence. |
| Charalampidou et al. [45] | Software traceability | Not defined to 2016 | 155 | ACM Digital Library, IEEE Xplore, Springer Link, Science Direct | Exploring the goals of existing software traceability approaches and the methods used for empirical evaluation |
| This SMS | Software traceability | 2000 to 2020 | 63 | ACM Digital Library, IEEE Xplore, Springer Link, Science Direct, Wiley InterScience, EI Compendex, ISI Web of Science | Analyzing the impact of traceability on software maintenance and evolution |

## 8 CONCLUSIONS

This work aims to investigate the impact of traceability on software maintenance and evolution. Systematic mapping study was employed to explore the available literature and finally 63 primary studies were included. The extracted data from the studies was used to answer the six RQs defined to achieve the goal of this SMS. The results show that 11 software maintenance and evolution activities can be supported by using traceability. Change management is the most frequently mentioned maintenance and evolution activity supported by the traceability practices. The results also reveal that there is a lack of industrial evidence to validate the impact of using traceability for maintenance and evolution activities. Moreover, the benefits and costs of using traceability during software maintenance and evolution phase have also been identified and classified across eight and four categories, respectively. 13 approaches and 32 tools were identified that support the use of traceability in software maintenance and evolution phase, while it is still a challenge to use these tools in industrial settings. Furthermore, it has also been determined that the quality of traceability links and the performance of using traceability approaches and tools are the two main challenges that hinder practitioners from employing the traceability practices in software maintenance and evolution activities.

The results of this SMS provide meaningful implications for both researchers and practitioners in the software engineering community. Strong industrial evidence is needed to assess the effectiveness of the proposed traceability approaches in maintenance and evolution. In addition, more effective methods shall be proposed to quantitatively measure the cost-benefit ratio of using traceability in the maintenance and evolution activities.

ACKNOWLEDGEMENTS

This work is partially sponsored by the National Key R&D Program of China with Grant No. 2018YFB1402800 and the National Natural Science Foundation of China (NSFC) under Grant No. 61972292.

APPENDIX A. SELECTED STUDIES

[S1] M. Mirakhorli and J. Cleland-Huang. 2011. A pattern system for tracing architectural concerns. In Proceedings of the 18th Conference on Pattern Languages of Programs (PLoP), ACM, 5: 1-10.

[S2] S. Ibrahim, N. B. Idris, M. Munro, and A. Deraman. 2005. A requirements traceability to support change impact analysis. Asian Journal of Information Technology, 4(4): 345-355.

[S3] K. Jaber, B. Sharif, and C. Liu. 2013. A study on the effect of traceability links in software maintenance. IEEE Access, 1: 726-741.

[S4] M. Mirakhorli, Y. Shin, J. Cleland-Huang, and M. Cinar. 2012. A tactic-centric approach for automating traceability of quality concerns. In Proceedings of the 34th International Conference on Software Engineering (ICSE), IEEE, 639-649.

[S5] A. V. Knethen. 2001. A trace model for system requirements changes on embedded systems. In Proceedings of the 4th International Workshop on Principles of Software Evolution (IWPSE), ACM, 17-26.

[S6] A. Bianchi, A. R. Fasolino, and G. Visaggio. 2000. An exploratory case study of the maintenance effectiveness of traceability models. In Proceedings the 8th International Workshop on Program Comprehension (IWPC), IEEE, 149-158.

[S7] J. L. de la Vara, M. Borg, K. Wnuk, and L. Moonen. 2016. An industrial survey of safety evidence change impact analysis practice. IEEE Transactions on Software Engineering, 42(12): 1095-1117.

[S8] R. Brcina and M. Riebisch. 2008. Architecting for evolvability by means of traceability and features. In Proceedings the 23rd IEEE/ACM International Conference on Automated Software Engineering-Workshops (ASE), IEEE, 72-81.

[S9] J. Maâzoun, N. Bouassida, and H. Ben-Abdallah. 2016. Change impact analysis for software product lines. Journal of King Saud University - Computer and Information Sciences, 28(4): 364-380.

[S10] M. Shahid and S. Ibrahim. 2016. Change impact analysis with a software traceability approach to support software maintenance. In Proceedings of the 13th International Bhurban Conference on Applied Sciences and Technology (IBCAST), IEEE, 391-396.

[S11] S. Gerdes, S. Lehnert, and M. Riebisch. 2014. Combining architectural design decisions and legacy system evolution. In Proceedings of the 8th European Conference on Software Architecture (ECSA), Springer LNCS, 50-57.

[S12] D. Dzvonyar, S. Krusche, R. Alkadhi, and B. Bruegge. 2016. Context-aware user feedback in continuous software evolution. In Proceedings of IEEE/ACM International Workshop on Continuous Software Evolution and Delivery (CSED), IEEE, 12-18.

[S13] P. Mäder and A. Egyed. 2015. Do developers benefit from requirements traceability when evolving and maintaining a software system?. Empirical Software Engineering, 20(2): 413-441.

[S14] M. Borg, O. C. Gotel, and K. Wnuk. 2013. Enabling traceability reuse for impact analyses: A feasibility study in a safety context. In Proceedings of the 7th International Workshop on Traceability in Emerging Forms of Software Engineering (TEFSE), IEEE, 72-78.

[S15] J. Cleland-Huang, C. K. Chang, and M. Christensen. 2003. Event-based traceability for managing evolutionary change. IEEE Transactions on Software Engineering, 29(9): 796-810.

[S16] M. Riebisch and I. Philippow. 2001. Evolution of product lines using traceability. In Proceeding of Workshop on Engineering Complex Object-Oriented Systems for Evolution (ECOOSE), DSG, 1-5.

[S17] J. Sametinger and M. Riebisch. 2002. Evolution support by homogeneously documenting patterns, aspects and traces. In Proceedings of the 6th European Conference on Software Maintenance and Reengineering (CSMR), IEEE, 134-140.

[S18] L. Passos, K. Czarnecki, S. Apel, A. Wąsowski, C. Kästner, and J. Guo. 2013. Feature-oriented software evolution. In Proceedings of the 7th International Workshop on Variability Modelling of Software-intensive Systems (VAMOS), ACM, 1-8.

[S19] L. Wen, D. Tuffley, and R. G. Dromey. 2014. Formalizing the transition from requirements' change to design change using an evolutionary traceability model. Innovations in Systems and Software Engineering, 10(3): 181-202.

[S20] A. Azmi and S. Ibrahim. 2011. Formulating a software traceability model for integrated test documentation: a case study. International Journal of Information and Electronics Engineering, 1(2): 178-184.

[S21] L. Naslavsky, H. Ziv, and D. J. Richardson. 2010. Mbsrt2: Model-based selective regression testing with traceability. In Proceedings of the 3rd International Conference on Software Testing, Verification and Validation (ICST), IEEE, 89-98.

[S22] G. Antoniol, E. Merlo, Y. G. Guéhéneuc, and H. Sahraoui. 2005. On feature traceability in object oriented programs. In Proceedings of the 3rd international workshop on Traceability in Emerging Forms of Software Engineering (TEFSE), ACM, 73-78.

[S23] Y. Zhang, R. Witte, J. Rilling, and V. Haarslev. 2008. Ontological approach for the semantic recovery of traceability links between software artefacts. IET Software, 2(3): 185-203.

[S24] N. Ubayashi, H. Akatoki, and J. Nomura. 2009. Pointcut-based architectural interface for bridging a gap between design and implementation. In Proceedings of the Workshop on AOP and Meta-Data for Software Evolution (RAM-SE), ACM, 5: 1-6.

[S25] A. Von Knethen and M. Grund. 2003. QuaTrace: a tool environment for (semi-) automatic impact analysis based on traces. In Proceedings of the 19th International Conference on Software Maintenance (ICSM), IEEE, 246-255.

[S26] M. Grechanik, K. S. McKinley, and D. E. Perry. 2007. Recovering and using use-case-diagram-to-source-code traceability links. In Proceedings of the 6th Joint Meeting of the European Software Engineering Conference and the ACM SIGSOFT symposium on The Foundations of Software Engineering (ESEC/FSE), ACM, 95-104.

[S27] G. Antoniol, G. Canfora, G. Casazza, A. De Lucia, and E. Merlo. 2002. Recovering traceability links between code and documentation. IEEE Transactions on Software Engineering, 28(10): 970-983.

[S28] Y. Li, J. Li, Y. Yang, and M. Li. 2008. Requirement-centric traceability for change impact analysis: a case study. In Proceedings of International Conference on Software Process (ICSP), Springer LNCS, 100-111.

[S29] J. H. Hayes, A. Dekhtyar, S. K. Sundaram, E. A. Holbrook, S. Vadlamudi, and A. April. 2007. REquirements TRacing On target (RETRO): improving software maintenance through traceability recovery. Innovations in Systems and Software Engineering, 3(3): 193-202.

[S30] Z. Mcharfi, B. El Asri, I. Dehmouch, A. Baya, and A. Kriouile. 2015. Return on investment of software product line traceability in the short, mid and long term. In Proceedings of the 17th International Conference on Enterprise Information Systems (ICEIS), SciTePress, 463-468.

[S31] W. Yu and S. Smith. 2009. Reusability of FEA software: a program family approach. In Proceedings of the 2nd International Workshop on Software Engineering for Computational Science and Engineering (SE-CSE), IEEE, 43-50.

[S32] S. Lehnert and M. Riebisch. 2013. Rule-based impact analysis for heterogeneous software artifacts. In Proceedings of the 17th European Conference on Software Maintenance and Reengineering (CSMR), IEEE, 209-218.

[S33] H. Omote, K. Sasaki, H. Kaiya, and K. Kaijiri. 2004. Software evolution support using traceability link between UML diagrams. In Proceedings of the 6th Joint Conference on Knowledge-Based Software Engineering (KBSE), IOS Press, 15-23.

[S34] M. Riebisch. 2004. Supporting evolutionary development by feature models and traceability links. In Proceedings of the 11th IEEE International Conference and Workshop on the Engineering of Computer-Based Systems (ECBS), IEEE, 370-377.

[S35] E. Ben Charrada, A. Koziolek, and M. Glinz. 2015. Supporting requirements update during software evolution. Journal of Software: Evolution and Process, 27(3): 166-194.

[S36] R. Settimi, J. Cleland-Huang, O. B. Khadra, J. Mody, W. Lukasik, and C. DePalma. 2004. Supporting software evolution through dynamically retrieving traces to UML artifacts. In Proceedings of the 7th International Workshop on Principles of Software Evolution (IWPSE), IEEE, 49-54.

[S37] M. A. Javed and U. Zdun. 2015. The supportive effect of traceability links in change impact analysis for evolving architectures - two controlled experiments. In Proceedings of the 14th International Conference on Software Reuse (ICSR), Springer LNCS, 139-155.

[S38] W. B. Santos, E. S. de Almeida, and S. R. D. L. Meira. 2012. TIRT: a traceability information retrieval tool for software product lines projects. In Proceedings of the 38th Euromicro Conference on Software Engineering and Advanced Applications (SEAA), IEEE, 93-100.

[S39] A. Ahmad, H. Basson, L. Deruelle, and M. Bouneffa. 2010. Towards an integrated quality-oriented modeling approach for software evolution control. In Proceedings of the 2nd International Conference on Software Technology and Engineering (ICSTE), IEEE, Volume 2, 320-324.

[S40] M. Konersmann, Z. Durdik, M. Goedicke, and R. H. Reussner. 2013. Towards architecture-centric evolution of long-living systems (the ADVERT approach). In Proceedings of the 9th International ACM SIGSOFT Conference on Quality of Software Architectures (QoSA), ACM, 163-168.

[S41] L. Naslavsky, H. Ziv, and D. J. Richardson. 2007. Towards leveraging model transformation to support model-based testing. In Proceedings of the 22nd IEEE/ACM International Conference on Automated Software Engineering (ASE), ACM, 509-512.

[S42] S. Winkler. 2009. Trace retrieval for evolving artifacts. In Proceedings of the 5th Workshop on Traceability in Emerging Forms of Software Engineering (TEFSE), IEEE, 49-56.

[S43] Y. Yu, J. Jurjens, and J. Mylopoulos. 2008. Traceability for the maintenance of secure software. In Proceedings of the 24th International Conference on Software Maintenance (ICSM), IEEE, 297-306.

[S44] A. Tang, P. Liang, V. Clerc, and H. Van Vliet. 2011. Traceability in the Co-evolution of Architectural Requirements and Design. In Relating Software Requirements and Architectures, Springer, 35-60.

[S45] M. Mirakhorli and J. Cleland-Huang. 2011. Using tactic traceability information models to reduce the risk of architectural degradation during system maintenance. In Proceedings of the 27th International Conference on Software Maintenance (ICSM), IEEE, 123-132.

[S46] L. Linsbauer, S. Fischer, R. E. Lopez-Herrejon, and A. Egyed. 2015. Using traceability for incremental construction and evolution of software product portfolios. In Proceedings of the 8th International Symposium on Software and Systems Traceability (SST), IEEE, 57-60.

[S47] S. A. Ajila and A. B. Kaba. 2004. Using traceability mechanisms to support software product line evolution. In Proceedings of the IEEE International Conference on Information Reuse and Integration (IRI), IEEE, 157-162.

[S48] S. Rochimah, W. M. N. Wan Kadir, and A. H. Abdullah. 2011. Utilizing multifaceted requirement traceability approach: a case study. International Journal of Software Engineering and Knowledge Engineering, 21(4): 571-603.

[S49] D. Kchaou, N. Bouassida, and H. Ben-Abdallah. 2017. A new approach for traceability between UML models. In Proceedings of the 12th International Conference on Software Technologies (ICSOFT), SciTePress, 128-139.

[S50] E. Bouillon, P. Mäder, and I. Philippow. 2013. A survey on usage scenarios for requirements traceability in practice. In Proceedings of the 19th International Working Conference on Requirements Engineering: Foundation for Software Quality (REFSQ), Springer LNCS, 158-173.

[S51] M. Rath, D. Lo, and P. Mäder. 2018. Analyzing requirements and traceability information to improve bug localization. In Proceedings of the 15th International Conference on Mining Software Repositories (MSR), ACM, 442-453

[S52] A. Goknil, I. Kurtev, and K. Van Den Berg. 2008. Change impact analysis based on formalization of trace relations for requirements. In ECMDA Traceability Workshop (ECMDA-TW), SINTEF Report, 59-75.

[S53] A. D. A. Gilberto Filho and A. Zisman. 2017. D3TraceView: A traceability visualization tool. In Proceedings of the 29th International Conference on Software Engineering and Knowledge Engineering (SEKE), KSI, 590-595

[S54] M. M. Rejab, N. F. M. Azmi, and S. Chuprat. 2019. Fuzzy Delphi Method for Evaluating HyTEE Model (Hybrid Software Change Management Tool with Test Effort Estimation). International Journal of Advanced Computer Science and Applications, 10(4): 529-535.

[S55] S. Charalampidou, A. Ampatzoglou, A. Chatzigeorgiou, and N. Tsiridis. 2018. Integrating traceability within the IDE to prevent requirements documentation debt. In Proceedings of the 44th Euromicro Conference on Software Engineering and Advanced Applications (SEAA), IEEE, 421-428.

[S56] Y. Zheng, C. Cu, and R. N. Taylor. 2018. Maintaining architecture-implementation conformance to support architecture centrality: From single system to product line development. ACM Transactions on Software Engineering and Methodology, 27(2): 1-52.

[S57] Y. Zheng, C. Cu, and H. U. Asuncion. 2017. Mapping features to source code through product line architecture: Traceability and conformance. In Proceedings of the IEEE International Conference on Software Architecture (ICSA), IEEE, 225-234.

[S58] P. Rempel and P. Mäder. 2016. Preventing defects: The impact of requirements traceability completeness on software quality. IEEE Transactions on Software Engineering, 43(8): 777-797.

[S59] D. Kchaou, N. Bouassida, M. Mefteh, and H. Ben-Abdallah. 2019. Recovering semantic traceability between requirements and design for change impact analysis. Innovations in Systems and Software Engineering, 15(2): 101-115.

[S60] I. D. Rubasinghe, D. A. Meedeniya, and I. Perera. 2017. Towards traceability management in continuous integration with SAT-analyzer. In Proceedings of the 3rd International Conference on Communication and Information Processing (ICCIP), ACM, 77-81.

[S61] M. Eyl, C. Reichmann, and K. Müller-Glaser. 2017. Traceability in a fine grained software configuration management system. In Proceedings of the International Conference on Software Quality (SWQD), Springer LNBIP, 15-29.

[S62] D. Kchaou, N. Bouassida, and Hanêne Ben-Abdallah. 2017. UML models change impact analysis using a text similarity technique. IET Software, 11(1): 27-37.

[S63] M. Goodrum, J. Cleland-Huang, R. Lutz, J. Cheng, and R. Metoyer. 2017. What Requirements Knowledge Do Developers Need to Manage Change in Safety-Critical Systems?. In Proceedings of the 25th International Requirements Engineering Conference (RE), IEEE, 90-99.

APPENDIX B. DISTRIBUTION OF SELECTED STUDIES OVER PUBLICATION VENUES

| # | Publication venue | Type | No. | % |
|---|---|---|---|---|
| 1 | Transactions on Software Engineering | Journal | 4 | 6.3 |
| 2 | Innovations in Systems and Software Engineering | Journal | 3 | 4.8 |
| 3 | IEEE International Conference on Software Maintenance and Evolution (ICSME) | Conference | 3 | 4.8 |
| 4 | IET Software | Journal | 2 | 3.2 |
| 5 | International Conference on Software Analysis, Evolution and Reengineering (SANER) | Conference | 2 | 3.2 |
| 6 | Euromicro Conference on Software Engineering and Advanced Applications (SEAA) | Conference | 2 | 3.2 |
| 7 | IEEE International Conference on Software Architecture (ICSA) | Conference | 2 | 3.2 |
| 8 | International Workshop on Traceability in Emerging Forms of Software Engineering (TEFSE) | Workshop | 2 | 3.2 |
| 9 | International Workshop on Principles of Software Evolution (IWPSE) | Workshop | 2 | 3.2 |
| 10 | ACM Transactions on Software Engineering and Methodology | Journal | 1 | 1.6 |
| 11 | Empirical Software Engineering | Journal | 1 | 1.6 |
| 12 | Journal of Software: Evolution and Process | Journal | 1 | 1.6 |
| 13 | International Journal of Software Engineering and Knowledge Engineering | Journal | 1 | 1.6 |
| 14 | IEEE Access | Journal | 1 | 1.6 |
| 15 | International Journal of Information and Electronics Engineering | Journal | 1 | 1.6 |
| 16 | International Journal of Advanced Computer Science and Applications | Journal | 1 | 1.6 |
| 17 | Journal of King Saud University-Computer and Information Sciences | Journal | 1 | 1.6 |
| 18 | Asian Journal of Information Tech | Journal | 1 | 1.6 |
| 19 | International Conference of Software Engineering (ICSE) | Conference | 1 | 1.6 |
| 20 | International Conference on Automated Software Engineering (ASE) | Conference | 1 | 1.6 |
| 21 | European Conference on Software Architecture (ECSA) | Conference | 1 | 1.6 |
| 22 | International Conference on Software Quality (SWQD) | Conference | 1 | 1.6 |
| 23 | International Conference on Software Reuse (ICSR) | Conference | 1 | 1.6 |
| 24 | IEEE International Conference on Information Reuse and Integration (IRI) | Conference | 1 | 1.6 |
| 25 | International Conference on Software Technologies (ICSOFT) | Conference | 1 | 1.6 |
| 26 | International Conference on Program Comprehension (ICPC) | Conference | 1 | 1.6 |
| 27 | International Working Conference on Requirements Engineering: Foundation for Software Quality (REFSQ) | Conference | 1 | 1.6 |
| 28 | Joint Meeting of the European Software Engineering Conference and the ACM SIGSOFT Symposium on the Foundations of Software Engineering (ESEC/FSE) | Conference | 1 | 1.6 |
| 29 | International Requirements Engineering Conference (RE) | Conference | 1 | 1.6 |
| 30 | International Conference on Software Testing, Verification and Validation (ICST) | Conference | 1 | 1.6 |
| 31 | International Joint Conference on Knowledge-Based Software Engineering (JCKBSE) | Conference | 1 | 1.6 |
| 32 | International Bhurban Conference on Applied Sciences and Technology (IBCAST) | Conference | 1 | 1.6 |
| 33 | International Conference and Workshop on the Engineering of Computer-Based Systems (ECBS) | Conference | 1 | 1.6 |
| 34 | International Conference on Automated Software Engineering – Workshops (ASEW) | Conference | 1 | 1.6 |
| 35 | International Conference on Communication and Information Processing (ICCIP) | Conference | 1 | 1.6 |
| 36 | International Conference on Enterprise Information Systems (ICEIS) | Conference | 1 | 1.6 |

| 37 | International Conference on Mining Software Repositories (MSR) | Conference | 1 | 1.6 |
| 38 | International Conference on Software Engineering and Knowledge Engineering (SEKE) | Conference | 1 | 1.6 |
| 39 | International Conference on Software Technology and Engineering (ICSTE) | Conference | 1 | 1.6 |
| 40 | International Conference on Software Process (ICSP) | Conference | 1 | 1.6 |
| 41 | International Conference on Pattern Languages of Programs (PLoP) | Conference | 1 | 1.6 |
| 42 | International Workshop on Traceability in Emerging Forms of Software Engineering (TEFSE) | Workshop | 1 | 1.6 |
| 43 | Workshop on Software Engineering for Computational Science and Engineering (SECSE) | Workshop | 1 | 1.6 |
| 44 | ECMDA Traceability Workshop (ECMDA-TW) | Workshop | 1 | 1.6 |
| 45 | Workshop on AOP and Meta-Data for Software Evolution (RAM-SE) | Workshop | 1 | 1.6 |
| 46 | International Symposium on Software and Systems Traceability (SST) | Workshop | 1 | 1.6 |
| 47 | International Workshop on Continuous Software Evolution and Delivery (CSED) | Workshop | 1 | 1.6 |
| 48 | International Workshop on Variability Modelling of Software-Intensive Systems (VaMoS) | Workshop | 1 | 1.6 |
| 49 | Workshop on Engineering Complex Object-Oriented Systems for Evolution (ECOOSE) | Workshop | 1 | 1.6 |
| 50 | Relating Software Requirements and Architectures | Book chapter | 1 | 1.6 |

REFERENCES

[1] K. H. Bennett and V. T. Rajlich. 2000. Software maintenance and evolution: a roadmap. In Proceedings of the Conference on the Future of Software Engineering (FOSE), ACM, 73-87.

[2] J. Radatz, A. Geraci, and F. Katki. 1990. IEEE standard glossary of software engineering terminology. IEEE Standard, 610.12-1990.

[3] P. Mäder and A. Egyed. 2015. Do developers benefit from requirements traceability when evolving and maintaining a software system?. Empirical Software Engineering, 20(2): 413-441.

[4] P. Mäder and A. Egyed. 2012. Assessing the effect of requirements traceability for software maintenance. In Proceedings of the 28th International Conference on Software Maintenance (ICSM), IEEE, 171-180.

[5] K. Jaber, B. Sharif, and C. Liu. 2013. A study on the effect of traceability links in software maintenance. IEEE Access, 1: 726-741.

[6] Y. Yu, J. Jurjens, and J. Mylopoulos. 2008. Traceability for the maintenance of secure software. In Proceedings of the 24th International Conference on Software Maintenance (ICSM), IEEE, 297-306.

[7] R. F. Lafet and M. Maia. 2011. An Empirical Assessment of the Use of Execution Traces in Software Maintenance. In Proceedings of the 25th Brazilian Symposium on Software Engineering (SBES), IEEE, 154-163.

[8] L. Chen, M. A. Babar, and H. Zhang. 2010. Towards an evidence-based understanding of electronic data source. In Proceedings of the 14th International Conference on Evaluation and Assessment in Software Engineering (EASE), ACM, 135-138.

[9] C. Wohlin. 2014. Guidelines for snowballing in systematic literature studies and a replication in software engineering. In Proceedings of the 18th International Conference on Evaluation and Assessment in Software Engineering (EASE), ACM, 38: 1-10.

[10] M. Rahimi. 2016. Trace link evolution across multiple software versions in safety-critical systems. In Proceedings of the 38th International Conference on Software Engineering (ICSE), ACM, 871-874.

[11] V. Basili, G. Caldiera, and D. Rombach. 1994. The Goal Question Metric Approach, Encyclopedia of Software Engineering. John Wiley & Sons, New York, NY.

[12] B. Ramesh and M. Jarke. 2001. Toward reference models for requirements traceability. IEEE Transactions on Software Engineering, 27(1): 58-93.

[13] R. Dömges and K. Pohl. 1998. Adapting traceability environments to project-specific needs. Communications of the ACM, 41(12): 54-62.

[14] B. G. Glaser and A. L. Strauss. 2017. Discovery of Grounded Theory: Strategies for Qualitative Research. Routledge.

[15] S. Adolph, W. Hall, and P. Kruchten. 2011. Using grounded theory to study the experience of software development. Empirical Software Engineering, 16(4): 487-513.

[16] S. A. Bohner. 1996. Impact analysis in the software change process: a year 2000 perspective. In Proceedings of the 12th International Conference on Software Maintenance (ICSM), IEEE, 42-51.

[17] A. von Mayrhauser and M. Vans A. 1995. Program comprehension during software maintenance and evolution. IEEE Computer, 28(8): 44-55.

[18] B. Kitchenham and S. Charters. 2007. Guidelines for performing systematic literature reviews in software engineering (Vol. 5). Technical Report, Ver. 2.3, EBSE Technical Report. EBSE.

[19] K. Petersen, S. Vakkalanka, and L. Kuzniarz. 2015. Guidelines for conducting systematic mapping studies in software engineering: An update. Information and Software Technology, 64: 1-18.

[20] M. A. Javed and U. Zdun. 2014. A systematic literature review of traceability approaches between software architecture and source code. In Proceedings of the 18th International Conference on Evaluation and Assessment in Software Engineering (EASE), ACM, 16: 1-10.

[21] M. Borg, P. Runeson, and A. Ardö. 2014. Recovering from a decade: a systematic mapping of information retrieval approaches to software traceability. Empirical Software Engineering, 19(6): 1565-1616.

[22] T. Vale, E.S. de Almeida, V. Alves, U. Kulesza, N. Niu, and R. de Lima. 2017. Software product lines traceability: A systematic mapping study. Information and Software Technology, 84: 1-18.

[23] Y. Yu, J. Jurjens, and J. Mylopoulos. 2008. Traceability for the maintenance of secure software. In Proceedings of the 24th International Conference on Software Maintenance (ICSM), IEEE, 297-306.

[24] L. M. Berlin. 1993. Beyond program understanding: a look at programming expertise in industry. In Proceedings of the 5th Workshop on Empirical Studies of Programmers (ESP), Ablex Publishing, 6-25.

[25] F. Shull, J. Singer, and D. I. Sjøberg. 2008. Guide to Advanced Empirical Software Engineering. Springer.

[26] IEEE Std. 1219-1998, 1998. Standard for Software Maintenance. Los Alamitos CA., USA: IEEE.

[27] M. Heindl and S. Biffl. 2005. A case study on value-based requirements tracing. In Proceedings of the 10th European Software Engineering Conference and the 13th ACM SIGSOFT International Symposium on Foundations of Software Engineering (ESEC/FSE), ACM, 60-69.

[28] M. M. Lehman. 1979. On understanding laws, evolution, and conservation in the large-program life cycle. Journal of Systems and Software, 1: 213-221.

[29] IEEE. 1998. IEEE Standard for Software Maintenance. ed.: Software Engineering Standards Committee of the IEEE Computer Society.

[30] P. Tripathy and K. Naik, 2014. Software Evolution and Maintenance: A Practitioner's Approach. John Wiley & Sons.

[31] O. Gotel, J. Cleland-Huang, J.H. Hayes, A. Zisman, A. Egyed, P. Grünbacher, and G. Antoniol. 2012, September. The quest for ubiquity: A roadmap for software and systems traceability research. In Proceedings of the 20th IEEE International Requirements Engineering Conference (RE), IEEE, 71-80.

[32] A. M. D. Duarte, D. Duarte, and M. Thiry. 2016, TraceBoK: Toward a software requirements traceability body of knowledge. In Proceedings of the 24th IEEE International Requirements Engineering Conference (RE), IEEE, 236-245.

[33] S. Nair, J. L. De La Vara, and S. Sen, 2013. A review of traceability research at the requirements engineering conference[re@21]. In Proceedings of the 21st IEEE International Requirements Engineering Conference (RE), IEEE, 222-229.

[34] J. Cleland-Huang, O. Gotel, and A. Zisman. 2012. Software and Systems Traceability. Heidelberg: Springer.

[35] R. Wieringa, N. Maiden, N. Mead, and C. Rolland. 2006. Requirements engineering paper classification and evaluation criteria: A proposal and a discussion. Requirements Engineering, 11(1): 102-107.

[36] C. Wohlin, P. Runeson, P. A. D. M. S. Neto, E. Engström, I. do Carmo Machado, and E.S. De Almeida. 2013. On the reliability of mapping studies in software engineering. Journal of Systems and Software, 86(10): 2594-2610.

[37] G. Spanoudakis and A. Zisman, 2005. Software Traceability: A Roadmap. In Handbook of Software Engineering and Knowledge Engineering: Vol 3: Recent Advances, 395-428.

[38] P. Heck and A. Zaidman, 2014. Horizontal traceability for just-in-time requirements: the case for open source feature requests. Journal of Software: Evolution and Process, 26(12): 1280-1296.

[39] H. Schwarz. 2009. Towards a comprehensive traceability approach in the context of software maintenance. In Proceedings of the 13th European Conference on Software Maintenance and Reengineering (CSMR), IEEE, 339-342.

[40] M. Fockel, J. Holtmann, and J. Meyer. 2012. Semi-automatic establishment and maintenance of valid traceability in automotive development processes. In Proceedings of the 2nd International Workshop on Software Engineering for Embedded Systems (SEES), IEEE, 37-43.

[41] A. D. Lucia, F. Fasano, R. Oliveto, and G. Tortora. 2007. Recovering traceability links in software artifact management systems using information retrieval methods. ACM Transactions on Software Engineering and Methodology, 16(4): 13:1-50.

[42] E. Ben Charrada, D. Caspar, C. Jeanneret, and M. Glinz. 2011. Towards a benchmark for traceability. In Proceedings of the 12th International Workshop on Principles of Software Evolution and the 7th Annual ERCIM Workshop on Software Evolution (IWPSE-EVOL), ACM, 21-30.

[43] B. J. Williams and J. C. Carver. 2010. Characterizing software architecture changes: a systematic review. Information and Software Technology, 52(1): 31-51.

[44] V. Alves, N. Niu, C. Alves, and G. Valença. 2010. Requirements engineering for software product lines: A systematic literature review. Information and Software Technology, 52(8): 806-820.

[45] S. Charalampidou, A. Ampatzoglou, E. Karountzos, and P. Avgeriou. 2020. Empirical studies on software traceability: A mapping study. Journal of Software: Evolution and Process, 33(2): 1-28.

[46] S. Jayatilleke and R. Lai, 2018. A systematic review of requirements change management. Information and Software Technology, 93: (163-185).

[47] M. A. Austin and M. H. Samadzadeh, 2005. Software comprehension/maintenance: An introductory course. In Proceedings of the 18th IEEE International Conference on Systems Engineering (ICSEng), IEEE, 414-419.

[48] H. Zhong and Z. Su. 2015. An empirical study on real bug fixes. In Proceedings of the 37th IEEE International Conference on Software Engineering (ICSE), ACM, 913-923.

[49] J. Van Gurp, S. Brinkkemper, and J. Bosch, 2005. Design preservation over subsequent releases of a software product: a case study of Baan ERP. Journal of Software Maintenance and Evolution: Research and Practice, 17(4): 277-306.

[50] M. Fowler, K. Beck, J. Brant, W. Opdyke, and D. Roberts, Refactoring: Improving the Design of Existing Code. Addison-Wesley Professional, 2018.

DATA AVAILABILITY STATEMENT

The data that support the findings of this study are available from the corresponding author upon reasonable request.